\documentclass[fleqn,usenatbib,useAMS]{mnras}

\usepackage{newtxtext,newtxmath}

\usepackage[T1]{fontenc}
\usepackage{ae,aecompl}

\usepackage{graphicx}	
\usepackage{amsmath}	
\usepackage{multicol}        
\usepackage{bm}		
\usepackage{pdflscape}	

\title[Origin of $r$-process-enhanced stars]{Origin of highly $r$-process-enhanced stars in a cosmological zoom-in simulation of a Milky Way-like galaxy}

\author[Y. Hirai et al.]{Yutaka Hirai$^{1, 2, 3}$\thanks{E-mail: yutaka.hirai@astr.tohoku.ac.jp}\thanks{JSPS Research Fellow}, 
Timothy C. Beers$^{1, 3}$, 
Masashi Chiba$^{2}$,
Wako Aoki$^{4}$,
Derek Shank$^{1, 3}$,
Takayuki R. Saitoh$^{5}$,\newauthor
Takashi Okamoto$^{6}$ and 
Junichiro Makino$^{5}$
\\
$^{1}$Department of Physics and Astronomy, University of Notre Dame, 225 Nieuwland Science Hall, Notre Dame, IN 46556, USA\\
$^{2}$Astronomical Institute, Tohoku University, 6-3 Aoba, Aramaki, Aoba-ku, Sendai, Miyagi 980-8578, Japan\\
$^{3}$Joint Institute for Nuclear Astrophysics, Center for the Evolution of the Elements (JINA-CEE), USA\\
$^{4}$National Astronomical Observatory of Japan, 2-21-1 Osawa, Mitaka, Tokyo 181-8588, Japan\\
$^{5}$Department of Planetology, Kobe University, 1-1 Rokkodai-cho, Nada-ku, Kobe, Hyogo 657-8501, Japan\\
$^{6}$Department of Physics, Faculty of Science, Hokakido University, N10 W8 Kita-ku, Sapporo, Hokkaido 060-0810, Japan
}

\date{Accepted 2022 August 30. Received 2022 August 28; in original form 2022 June 08}

\pubyear{2022}

\begin{document}
\label{firstpage}
\pagerange{\pageref{firstpage}--\pageref{lastpage}}
\maketitle

\begin{abstract}
The $r$-process-enhanced (RPE) stars provide fossil records of the assembly history of the Milky Way and the nucleosynthesis of {the heaviest elements}. Observations by the $R$-Process Alliance (RPA) and others have confirmed that many RPE stars are associated with chemo-dynamically tagged groups, which likely came from accreted dwarf galaxies of the Milky Way (MW). However, we do not know how RPE stars are formed. Here, we {present the result of a cosmological zoom-in simulation of an MW-like galaxy with $r$-process enrichment, performed with the highest resolution in both time and mass. Thanks to this advancement, unlike previous simulations, we find that most highly RPE ($r$-II; [Eu/Fe] $> +0.7$) stars are formed in low-mass dwarf galaxies that have been enriched in $r$-process elements for [Fe/H] $\,<-2.5$}, while those with higher metallicity are formed \textit{in situ}, in locally enhanced gas clumps that were not necessarily members of dwarf galaxies. This result suggests that low-mass accreted dwarf galaxies are the main formation site of $r$-II stars with [Fe/H] $\,<-2.5$. We also find that most low-metallicity $r$-II stars exhibit halo-like kinematics. Some $r$-II stars formed in the same halo show low dispersions in [Fe/H] and somewhat larger dispersions of [Eu/Fe], similar to the observations. The fraction of simulated $r$-II stars is commensurate with observations from the RPA, and the distribution of the predicted [Eu/Fe] for halo $r$-II stars matches that observed. These results demonstrate that RPE stars can be valuable probes of the {accretion of dwarf galaxies in the early stages of their formation}.
\end{abstract}




\section{Introduction}\label{sec:intro}

The chemical abundances of stars potentially provide clues to understanding galaxy formation and nucleosynthesis. In particular, elements synthesized by the rapid neutron-capture process ($r$-process elements), such as Eu, Th, and U, are crucial elements for resolving early Galactic chemical evolution \citep[e.g.][]{2005ARA&A..43..531B, 2008ARA&A..46..241S, 2015ARA&A..53..631F, 2017ARNPS..67..253T, 2018ARNPS..68..237F, 2021RvMP...93a5002C}, using low-metallicity stellar probes in the Milky Way (MW) \citep[e.g.][]{1978A&A....67...23S, 1981A&A....97..391T, 1997ARA&A..35..503M}.

Recognition of the  highly $r$-process-enhanced (RPE) star CS 22892-052 (\citealt{1994ApJ...431L..27S}; [Eu/Fe]\footnote{[A/B] = $\log_{10}({N_{\mathrm{A}}}/{N_{\mathrm{B}}})-\log_{10}({N_{\mathrm{A}}}/{N_{\mathrm{B}}})_{\odot}$, where $N_{\mathrm{A}}$ and $N_{\mathrm{B}}$ are the number densities of elements A and B, respectively.} $= +1.6$ at [Fe/H] $= -3.1$), discovered in the HK survey \citep{1985AJ.....90.2089B, 1992AJ....103.1987B}, made it clear that the abundances of $r$-process elements in the early epochs of star formation in the MW were not homogeneous. 
Later observations of this star have been reported by many authors, including \citet{1996ApJ...467..819S}, \citet{2003ApJ...591..936S}, and \citet{2005A&A...439..129B}. 

The general definition of RPE stars are those exhibiting enhancement of heavy $r$-process elements thought to be primarily produced by the $r$-process such as europium (Eu; $Z =$ 63), [Eu/Fe]$\,>\,+0.3$. 
Subsequent to the recognition of CS 22892-052, $r$-process-element abundance measurements for additional RPE stars were reported: e.g. CS 31082-001 \citep{2001Natur.409..691C, 2002A&A...387..560H, 2004A&A...428L...9P}, CS 22183-031 \citep{2004ApJ...607..474H}, and HE 1523-0901 \citep{2007ApJ...660L.117F}, among many others. Furthermore, a  star-to-star scatter of [Eu/Fe] (and [Sr/Fe]) ratios (more than two orders of magnitude) has been confirmed for stars with [Fe/H]$\,<-$2.5 \citep[e.g. ][]{1995AJ....109.2757M, 1996ApJ...471..254R, 2007A&A...476..935F, 2013ApJ...771...67I, 2014AJ....147..136R}. Most recently, ground and space-based observations of the bright metal-poor ($V = 9$, [Fe/H] $=-1.46$) highly RPE ([Eu/Fe] $= +1.32$) horizontal-branch star HD~225925 has produced the largest inventory of measured elemental abundances for any star other than the Sun \citep{roederer2022}, and provided a new `template star' for the observed distribution of $r$-process elements.

\subsection {The classification of RPE stars and their observed environments}

RPE stars are conventionally sub-classified by their [Eu/Fe] ratios, originally defined by \citet{2005ARA&A..43..531B} in order to separate moderately and highly enhanced stars ($r$-I: +0.3$\,<\,$[Eu/Fe]$\,\leq\,$+1.0, [Ba/Eu]$\,<\,$0; $r$-II [Eu/Fe]$\,>\,$+1.0, [Ba/Eu]$\,<\,$0). The condition [Ba/Eu]$\,<\,$0 is used in order to select stars dominantly enriched by the $r$-process, rather than the $s$-process\footnote{According to \citet{2000ApJ...544..302B}, the pure $s$- and $r$-process ratios of [Ba/Eu] are [Ba/Eu]$_{s}\,=\,+1.5$ and [Ba/Eu]$_{r}\,= -0.82$, respectively.}. \citet{2020ApJS..249...30H} recently statistically reclassified RPE stars, based on the large sample of  such stars reported by the RPA. They defined the $r$-I and $r$-II boundaries, using a $k$-medoids partitioning technique \citep{1990fgda.book.....K} and a mixture-model analysis. Their revised classification of $r$-I and $r$-II stars is:\\

$r$-I: +0.3$\,<\,$[Eu/Fe]$\,\leq\,$+0.7, [Ba/Eu]$\,<\,$0,

$r$-II: [Eu/Fe]$\,>\,$+0.7, [Ba/Eu]$\,<\,$0.\\

\noindent This classification was defined to minimize the distance between the members of three groups (the third group being those stars with [Eu/Fe] $\le +0.3$) to determine the group centres in the [Eu/Fe] distribution. They also found that stars classified into these three groups more appropriately represented the sample [Eu/Fe] distribution than those classified into two groups, based on the Akaike information criterion \citep{10.1093/biomet/60.2.255}.
Recently, \citet{2020ApJ...898...40C} defined $r$-III stars for the handful of stars with [Eu/Fe]$\,>\,$+2.0 and [Ba/Eu]$\,<-0.5$. 

Since the $r$-process-element abundances of $r$-II stars are well above the average values in the MW, they are ideal probes to investigate their formation environment. As we point out below, the formation of $r$-I stars in our simulation is heavily affected by the assumption of the appropriate yields, which is highly uncertain. Thus, in this paper, we concentrate on the $r$-II stars (for simplicity, we combine these stars with the few $r$-III stars known and simply refer to their union as `$r$-II' stars). The $r$-I stars will be considered in detail in a forthcoming paper.

RPE stars are also found in dwarf galaxies. The ultrafaint dwarf (UFD) galaxy Reticulum II is a notable example \citep{2016Natur.531..610J, 2016ApJ...830...93J, 2016AJ....151...82R, 2022arXiv220703499J}. Seven out of nine observed stars are classified as $r$-II stars in this galaxy. \citet{2016Natur.531..610J} argued that a nucleosynthetic event with a europium yield of $10^{-4.5}\,$M$_{\sun}$ could enhance the $r$-process-element abundances of gas with a mass $\sim$10$^{6}\,$M$_{\sun}$ to the level required for $r$-II stars. The Tucana III UFD is another example. An $r$-I star, DES J235532.66-593114.9, with [Eu/Fe] = +0.60, is confirmed in this galaxy, along with at least two other less-enhanced $r$-I stars \citep{2017ApJ...838...44H, 2019ApJ...882..177M}. Several RPE stars are also found in more massive dwarf galaxies: Fornax \citep{2010A&A...523A..17L, 2021ApJ...912..157R}, Ursa Minor \citep{2001ApJ...548..592S, 2004PASJ...56.1041S}, Carina \citep{2003AJ....125..684S, 2017ApJS..230...28N}, Draco \citep{2009ApJ...701.1053C}, Sculptor \citep{2003AJ....125..684S, 2005AJ....129.1428G} dwarf spheroidal galaxies (dSphs) and the Large/Small Magellanic Clouds \citep[LMC/SMC;][]{2021AJ....162..229R}. These findings clearly suggest a link between the formation of the MW and dwarf galaxies of a variety of masses.

\subsection{The nature of the stellar halo of the MW}

Studies of the nature of the MW halo have a long, rich history (see, e.g. \citealt{2005ARA&A..43..531B} for a brief overview). Two decades ago,  \citet{2000AJ....119.2843C} analysed the motion of $\sim$1200 stars with [Fe/H]$\,<-0.6$, based on the kinematically unbiased sample of over 2000 metal-poor stars from \citet{2000AJ....119.2866B}, and found that there is no correlation between stellar metallicities and their orbital eccentricities. This finding clearly indicated, for the first time, that the correlation seen in \citet{1962ApJ...136..748E} that led to a monolithic collapse scenario for MW formation was due to the kinematic selection bias employed to assemble their sample. Chiba \& Beers concluded that the halo was complex and not a single structure with a simple formation history, initiating the discussion of the possible existence of a flattened `inner halo' and more spherical `outer halo' with differing formation mechanisms.  

The series of papers by \citet{2007Natur.450.1020C,2010ApJ...712..692C} and \citet{2012ApJ...746...34B} amplified these conclusions based on the much larger sample of metal-poor stars obtained during the Sloan Digital Sky Survey \citep{2000AJ....120.1579Y}, in particular the stellar-specific sub-survey Sloan Extension for Galactic Understanding and Exploration \citep[SEGUE;][]{2009AJ....137.4377Y}, and clearly demonstrated `the halo' was, in fact, a superposition of at least two primary components -- an inner and an outer halo -- that exhibit different spatial density profiles, stellar orbits, and stellar metallicities. These properties indicate that the individual halo components likely formed in fundamentally different ways, through successive dissipational and dissipationless mergers and the tidal disruption of proto-Galactic clumps, refining the original suggestions from \citet{2000AJ....119.2843C}.

In the context of this scenario, stars in the MW can be broadly classified as members of \textit{in situ} (primarily inner-halo) and accreted (primarily outer-halo) components. \textit{In situ} stars are thought to have formed in what is sometimes referred to as the `main halo' of the MW (dominated by the inner-halo population), while accreted stars are expected to have formed in the halo outside of the MW's main halo and accreted by the MW at a later time, and are expected to predominantly populate the outer halo.  

\subsection{Theoretical expectations and halo substructure}
According to the hierarchical structure formation scenario in the Lambda Cold Dark Matter ($\Lambda$CDM) paradigm, galaxies could be formed by the clustering of smaller systems \citep[e.g.][]{1978MNRAS.183..341W, 1984Natur.311..517B}. Therefore, it is crucial to find signatures of the accretion events of dwarf galaxies to the MW. Relaxation of the initial dynamical phase-space distribution is thought to exceed the age of the Universe in order to fully mix the stellar orbital energies and angular momenta. Thus, the observed phase-space distribution of stars has been used to identify substructures in the MW \citep[and references therein]{2020ARA&A..58..205H}. From a sufficiently large sample of stars with available dynamical information, the assembly of the MW can be reconstructed. If stellar metallicities are also available, one can use this to make a rough inference of the ages of the stars involved. Results based on precision age-determination techniques (in particular, for subgiant stars) have been recently reported for halo and disc-system stars by \citet{2022Natur.603..599X}.  

The astrometric data provided by the \textit{Gaia} satellite mission has already revolutionized our understanding of the building blocks of the MW \citep{2016A&A...595A...1G, 2016A&A...595A...2G, 2018A&A...616A...1G, 2021A&A...649A...1G}. Several substructures have been identified in the dynamical phase space, which could be remnants of disrupted dwarf galaxies accreted to the MW \citep[e.g.][]{2018MNRAS.475.1537M, 2018ApJ...860L..11K, 2019A&A...631L...9K, 2018MNRAS.478..611B, 2018ApJ...863..113H, 2018Natur.563...85H, 2020ApJ...891...39Y, 2020ApJ...901...48N}. Chemical-abundance analyses of member stars in each substructure show that some of these have chemical abundances similar to presently observed MW satellites \citep[e.g.][]{2019ApJ...874L..35M, 2021A&A...650A.110M, 2022arXiv220311808M, 2021ApJ...908L...8A, 2021ApJ...913L..28L, 2022MNRAS.513.1557C, 2022arXiv220404233H}. Numerous stellar debris streams have also been reported (e.g. \citealt{2022arXiv220410326M}); their full chemical compositions are only just now beginning to be studied. Notably, \citet{2021ApJ...912...52G} have reported the detection of stars that exhibit clear $r$-process patterns in several streams, including a number of RPE stars. 

Statistical analyses based on the combination of spectroscopic and \textit{Gaia} astrometric data for a variety of stellar samples has been used to identify Dynamically Tagged Groups (DTGs) in the phase-space distribution \citep{2020ApJ...891...39Y, 2021ApJ...907...10L, 2022ApJ...926...26S, 2022ApJS..261...19S}. In the largest such sample studied to date, \citet{2022ApJS..261...19S} applied an unsupervised learning algorithm Hierarchical Density-Based Spatial Clustering of Applications with Noise \citep[HDBSCAN;][]{10.1007/978-3-642-37456-2_14} for orbital energy and cylindrical actions to some 8000 stars selected from the RAdial Velocity Experiment (RAVE) DR6 \citep{2006AJ....132.1645S, 2020AJ....160...82S, 2020AJ....160...83S};  they identified 179 DTGs. Most of these DTGs are associated with known substructures such as Gaia-Sausage-Enceladus \citep[GSE,][]{2018MNRAS.478..611B, 2018Natur.563...85H}, the Metal-Weak Thick Disc \citep{1990AJ....100.1191M, 2014ApJ...794...58B, 2019ApJ...887...22C}, the Splashed Disc \citep{2019A&A...632A...4D, 2020MNRAS.494.3880B}, Thamnos \citep{2019A&A...631L...9K}, the Helmi Stream \citep{1999Natur.402...53H, 2000AJ....119.2843C}, LMS-1/Wukong \citep{2020ApJ...891...39Y, 2020ApJ...901...48N}. 
They also found 22 DTGs that are associated with RPE stars.

\subsection{RPE stars as probes of halo substructure}

In a seminal investigation, \citet{2018AJ....156..179R} showed that RPE stars tend to be associated with groups of stars with similar orbits, based on an analysis of the dynamics of 35 $r$-II stars using \textit{Gaia} DR2 \citep{2018A&A...616A...1G} and previously published radial velocities and elemental abundances \citep[see also][]{2022arXiv220704110H}. They identified eight groups of 2--4 stars each, with an indication that the stars within each group had more similar abundances of [Fe/H] and [Eu/Fe] than would have been expected from random draws. Their orbital analysis also showed that their sample stars all exhibited halo-like kinematics. These results suggested that RPE stars may have arisen from disrupted dwarf galaxies similar to UFDs or low-luminosity dSphs.

The ongoing $R$-Process Alliance (RPA) effort has greatly increased the number of confirmed RPE stars \citep{2018ApJ...858...92H, 2018ApJ...868..110S, 2020ApJ...898..150E, 2020ApJS..249...30H}. \citet{2020ApJS..249...30H} reported a total of 72 $r$-II stars and 232 $r$-I stars based on this survey alone. \citet{2020ApJsubGudin} analysed the dynamics of 446 RPE stars in the RPA and additional literature samples, again using HDBSCAN, to identify 30 dynamical groups of RPE stars with between 3 and 12 members each. Since they performed the clustering analysis exclusively for chemically peculiar stars (RPE stars), they call these groups Chemo-Dynamically Tagged Groups (CDTGs). They found that each CDTG has smaller than expected dispersions in metallicity ([Fe/H]), carbonicity ([C/Fe]), and neutron-capture-element abundance ratios ([Sr/Fe], [Ba/Fe], and [Eu/Fe]). These results strongly suggested that RPE stars within each CDTG experienced a common chemical evolution, supporting the scenario that RPE stars in CDTGs were from disrupted dwarf galaxies or globular clusters. Most recently, \citet{shank2022c} have compiled a sample of $\sim$1800 RPE stars based on the RPA, GALAH DR3 \citep{2021MNRAS.506..150B}, and other literature sources and described the abundance dispersion results for 38 CDTGs with between 5 and 20 members each.
\subsection{The origin of RPE stars and their birth environments}

The enrichment of $r$-process elements in the MW and dwarf galaxies with chemical-evolution models has been studied for over thirty years \cite[e.g.][]{1990Natur.345..491M, 1992ApJ...391..719M, 1999ApJ...511L..33I,2000ApJ...531L..33T, 2004A&A...416..997A, 2004ApJ...600L..47I, 2015ApJ...804L..35I, 2014MNRAS.438.2177M,2014ApJ...783..132K, 2014A&A...565A..51C, 2014A&A...565L...5T, 2015A&A...577A.139C, 2015MNRAS.452.1970W, 2015ApJ...814...41H, 2017MNRAS.466.2474H, 2018ApJ...855...99C, 2019ApJ...875..106C, 2018ApJ...865...87O, 2019ApJ...876...28S, 2019MNRAS.487..580S, 2021MNRASW}. Most of these studies focused on constraining the astrophysical sites of the $r$-process elements by explaining the trend and dispersions of [Eu/Fe] ratios as a function of [Fe/H]. \citet{2004A&A...416..997A} preferred core-collapse supernovae (CCSNe) as the astrophysical site rather than binary neutron star mergers (NSMs) in terms of the rates and delay times. On the other hand, \citet{2015ApJ...814...41H} have shown that NSMs can be a major contributor of $r$-process elements in dwarf galaxies because of their slow chemical evolution. Evidence for the (slightly) delayed production of the $r$-process elements, which would come from NSMs, has been seen in dwarf galaxies and accreted components \citep{2018ApJ...869...50D, 2020A&A...634L...2S, 2021A&A...650A.110M, 2022ApJ...926L..36N}.

Nucleosynthetically, NSMs are one of the most promising candidates for the astrophysical sites of the $r$-process \citep[e.g.][]{1974ApJ...192L.145L, 1982ApL....22..143S, 1989Natur.340..126E, 1989ApJ...343..254M, 1999ApJ...525L.121F, 2011ApJ...738L..32G,2015MNRAS.452.3894G, 2012MNRAS.426.1940K,2013ApJ...773...78B, 2014ApJ...789L..39W, 2015PhRvD..91f4059S, 2016PhRvD..93l4046S, 2016MNRAS.460.3255R, 2022arXiv220505557F}. For other sites, CCSNe, driven by magneto-rotational instability \citep{2012ApJ...750L..22W, 2015ApJ...810..109N, 2017ApJ...836L..21N} and collapsars \citep{2019Natur.569..241S} are recently proposed, while CCSNe driven by neutrino winds have been found to have difficulty synthesizing elements heavier than $A \sim$ 110 \citep{2011ApJ...726L..15W, 2018ApJ...852...40W, 2013ApJ...770L..22W}. Although contributions from each site to the $r$-process enrichment are still debated, many nucleosynthesis and chemical-evolution studies support that NSMs play an important role as the astrophysical site of the $r$-process. Observations of the kilonova associated with the gravitational wave event GW$170817$ have shown a definitive astrophysical source of heavy elements created by the $r$-process in NSMs \citep[e.g.][]{2017PhRvL.119p1101A, 2017ApJ...848L..12A, 2017Natur.551...64A, 2017Sci...358.1570D,2017Sci...358.1574S,  2017ApJ...848L..18N,  2017PASJ...69..102T, 2019Natur.574..497W, 2021ApJ...913...26D}.

Prolific amounts of $r$-process elements from NSMs could form RPE stars in UFDs  \citep{2017MNRAS.471.2088S, 2020MNRAS.494..120T, 2021MNRAS.506.1850J}. Chemical-evolution models following the hierarchical structure formation scenario or cosmological simulations implied that RPE stars in the MW are formed in UFD-size galaxies \citep{2018ApJ...865...87O, 2019ApJ...871..247B, 2021MNRASW}. However, the mechanism and environments of RPE star formation in the MW are not yet well-understood due to the lack of detailed simulations. In order to understand the formation of RPE stars in the MW, it is necessary to perform high-resolution cosmological zoom-in simulations that can resolve dwarf galaxies with stellar masses 
down to $\lesssim 10^5\,$M$_{\sun}$ with models of NSMs, metal diffusion, and sufficient numbers of snapshots to follow RPE star formation. Although sophisticated galaxy-formation simulations have been published recently \citep[e.g.][]{2011ApJ...742...76G, 2016ApJ...827L..23W, 2016MNRAS.457.1931S,2017MNRAS.467..179G, 2020MNRAS.491.3461B,2020MNRAS.498.1765F, 2021ApJ...906...96A, 2021MNRAS.503.5826A}, it is still challenging to perform simulations considering all of these requirements due to the high computational and storage costs.

{This study has performed the highest stellar mass and time resolution simulation of an MW-like galaxy with enrichment of $r$-process elements, which is able to distinguish between $r$-process enrichment in the UFD-like low-mass building blocks and \textit{in situ} (local) enrichment. Thanks to these advancements, we can study the kinematics of the simulated stars in order to compare with observed CDTGs.} 

{Our paper is outlined as follows. In Section \ref{sec:sim}, we summarize and compare previous cosmological simulations of MW-like galaxies with $r$-process enrichment with our study.} In Section \ref{sec:method}, we describe our code, adopted models, and initial conditions. Section \ref{sec:results} describes the chemical abundances and kinematics of our simulated stars. Section \ref{sec:discussion} considers the formation and chemo-dynamical properties of $r$-II stars. Section \ref{sec:conclusion} summarizes our conclusions and provides some perspectives on future efforts.
\section{Cosmological simulations of Milky Way-like galaxies with $r$-process enrichments}\label{sec:sim}
{Several previous studies have performed cosmological simulations of MW-like galaxies with $r$-process enrichment \citep{2015ApJ...807..115S, 2015MNRAS.447..140V, 2020MNRAS.494.4867V, 2022MNRAS.512.5258V, 2018MNRAS.477.1206N, 2019MNRAS.483.5123H}. \citet{2015ApJ...807..115S} investigated the enrichment of $r$-process elements in the Eris simulation \citep{2011ApJ...742...76G} using a post-processing model for NSMs, and showed that NSMs can explain the level of Eu abundance of metal-poor stars in an MW-like galaxy. \citet{2015MNRAS.447..140V} computed the enrichment of $r$-process elements in the FIRE simulation \citep{2014MNRAS.445..581H}. Their results suggested that large-scale hydrodynamic mixing processes play an essential role in explaining the observed scatter of [$r$-process/Fe] abundance ratios. \citet{2018MNRAS.477.1206N} discussed the distribution of Eu in MW-like galaxies formed in the IllustrisTNG simulations \citep{2018MNRAS.475..676S}, and found that the scatter of [Eu/Fe] is sensitive to gas properties in the redshift range $z$ = 2--4.}

{These studies mainly focused on understanding how to reproduce the observed scatter of [Eu/Fe] in low-metallicity stars. However, it has proven difficult to detail the formation mechanism of stars with $r$-process elements with these simulations. As described in Section \ref{sec:intro}, it is important to resolve UFD-size galaxies in cosmological simulations of MW-like galaxies. In addition, as shown in \citet{2021MNRAS.506.1850J}, it is necessary to resolve $r$-process enrichment with a timescale of $\sim$10 Myr to clarify the formation history of RPE stars. To date, no simulations have had sufficient mass and time resolution to discuss RPE star formation in the context of the formation and assembly history of the MW.}

{The main aims of this study are to (1) understand the mechanism and environments of RPE star formation and (2) provide a connection between the chemo-dynamical properties of RPE stars and the assembly history of the MW. Here we compute the enrichment of $r$-process elements in the highest stellar mass and time resolution cosmological zoom-in simulation of an MW-like galaxy, which can resolve dwarf galaxies down to stellar masses of $\sim$10$^5\,$M$_{\sun}$, with a snapshot interval of $\sim$10\,Myr. This simulation, with in total 42 TB outputs, makes it possible to extract information about RPE star formation and chemo-dynamical properties. As noted above, we focus on $r$-II stars ([Eu/Fe]$\,>\,+0.7$ and [Ba/Eu]$\,<\,0$) in the solar neighbourhood, defined in Section \ref{sec:solar}, and regard europium (Eu) as representative of the $r$-process elements in order to compare with observations.}

\section{Method and Models}\label{sec:method}

\subsection{Code}\label{sec:Code}

Here we briefly describe the code employed in this study. Details of the code are described in \citet{2008PASJ...60..667S} and \citet{2018ApJ...855...63H, 2019ApJ...885...33H}. 
In this study, we adopt the $N$-body/smoothed particle hydrodynamics (SPH) code \textsc{asura} \citep{2008PASJ...60..667S, 2009PASJ...61..481S}. Gravity is computed with a tree method \citep{1986Natur.324..446B} parallelized following \citet{2004PASJ...56..521M}. Hydrodynamics are computed using the density-independent SPH (DISPH) method, which can correctly treat instabilities in contact discontinuity \citep{2013ApJ...768...44S, 2016ApJ...823..144S}. A Fully Asynchronous Split Time-Integrator  (FAST) for a Self-Gravitating Fluid scheme is adopted to reduce the computational cost of the self-gravitating fluid systems in SPH \citep{2010PASJ...62..301S}. A time-step limiter, which enforces the time-step difference among local particles small enough to solve the propagation of shocks, is also implemented \citep{2009ApJ...697L..99S}. We adopt the metallicity dependent cooling and heating functions from 10 to 10$^9$ K generated by \textsc{cloudy} version 13.5 \citep{1998PASP..110..761F, 2013RMxAA..49..137F, 2017RMxAA..53..385F}. Effects of self-shielding \citep{2013MNRAS.430.2427R} and the ultra-violet background radiation field \citep{2012ApJ...746..125H} are also implemented in our simulations.

Star particles are stochastically created following the Schmidt law \citep{1959ApJ...129..243S}. Here we implement the models for star formation described in \citet{2003MNRAS.345..429O} 
and \cite{2008PASJ...60..667S}. In this model, gas particles need to satisfy conditions for star formation as follows: (1) the number density is higher than 100 cm$^{-3}$, (2) the temperature is lower than 1000 K, (3) the divergence of the velocity is less than zero, and (4) the particle is not heated by supernovae. The threshold density for star formation is similar to the mean density of giant molecular clouds \citep[e.g.][]{1987ApJ...319..730S, 2009ApJ...699.1092H}. 

When a gas particle satisfies all of these conditions, we compare a randomly generated number ($\mathcal{R}$) between 0 to 1 with the probability ($p_{*}$) in a given timestep ($\Delta{t}$): 
\begin{equation}
p_{*} = \frac{m_{\mathrm{gas}}}{m_{*}}
\left\{1-\exp\left(-c_{*}
\frac{\Delta t}{t_{\mathrm{dyn}}}\right)\right\},\label{eq:sf_probability}
\end{equation}
where $m_{\mathrm{gas}}$, $m_{*}$, $t_{\mathrm{dyn}}$, and $c_{*}$ are the mass of one gas particle, the mass of one star particle, the dynamical time of the star-forming region, and a dimensionless star-formation efficiency. Here we set $c_{*}$ = 0.5, motivated by observations of star-formation efficiency in star clusters \citep[e.g.][]{2003ARA&A..41...57L} and previous simulations \citep[e.g.][]{2006ApJ...641..878T, 2008ApJ...673..810T, 2008PASJ...60..667S}. If $\mathcal{R}\,<\,p_{*}$, one-third of the mass of gas particles are converted into star particles, {following \citet{2003MNRAS.345..429O,2005MNRAS.363.1299O} and \citet{2008PASJ...60..667S}, in order to prevent an abrupt decrease of the number of cold gas particles due to star formation.}

Star particles are approximated as simple stellar populations (SSPs). We adopt the initial mass function (IMF) of \citet{2003PASP..115..763C} from 0.1 to 100 M$_{\sun}$. The {metallicity-dependent} stellar lifetime table is taken from \citet{1998A&A...334..505P}. {Since mega metal-poor ([Fe/H]$\,<\,-6$) stars are mainly formed by molecular hydrogen cooling, rather than the fine-structure line cooling by heavy elements \citep{2005ApJ...626..627O}, Population III star-formation simulations predict top-heavy IMFs \citep[e.g.][]{2015MNRAS.448..568H, 2016MNRAS.462.1307S}. Therefore, we adopt the top-heavy IMF from 0.6 to 300 M$_{\sun}$ estimated from simulations of Population III star formation \citep{2014ApJ...792...32S} for stars with $Z\,<\,10^{-5}\,Z_{\sun}$. A larger number of low-mass stars are predicted to be formed towards higher metallicity for $Z\,\gtrsim\,10^{-5}\,Z_{\sun}$ \citep{2021MNRAS.508.4175C}.} The number of Lyman-$\alpha$ photons is evaluated with \textsc{p\'egase} \citep{1997A&A...326..950F} in order to heat gas around massive stars with ages less than 10 Myr to 10$^4$\,K as an H{\sc ii} region. The Chemical Evolution Library (\textsc{celib}) is adopted for the implementation of different nucleosynthetic sources \citep{2016ascl.soft12016S, 2017AJ....153...85S, 2021PASJ...73.1036H}. Stars from 13 to 40 M$_{\sun}$ for $Z\,\geq\,10^{-5}\,Z_{\sun}$ (from 13 to 300 M$_{\sun}$ for $Z\,<\,10^{-5}\,Z_{\sun}$) are assumed to explode as CCSNe. We adopt the yields from \citet{2013ARA&A..51..457N}. We also adopt a model of type Ia supernovae (SNe Ia), assuming a power-law delay-time distribution with an index of $-$1 and a minimum delay time of 40 Myr \citep[e.g.][]{2008PASJ...60.1327T, 2012PASA...29..447M, 2012MNRAS.426.3282M}. We adopt the yields of SNe Ia from the N100 model of \citet{2013MNRAS.429.1156S}. The minimum delay time for SNe Ia is taken from the lifetime of the likely most massive progenitors (8 M$_{\sun}$) of white dwarfs. We note that the short end of the delay-time distribution of SNe Ia is highly uncertain \citep{2020ApJ...890..140S}. Several chemical-evolution studies prefer a minimum delay time of SNe Ia longer than 500 Myr \citep{2015ApJ...799..230H,2021MNRASW}.

We assume that Eu is synthesized exclusively by NSMs. We adopt a model of NSMs with a delay-time distribution with power-law index $-$1, with a minimum delay time of 10 Myr, following population-synthesis calculations \citep[e.g.][]{2012ApJ...759...52D}. Recent studies have shown that the power-law index of $-$1 in NSMs cannot reproduce the decreasing trend of [Eu/Fe] towards higher metallicity seen in the MW disc \citep{2018IJMPD..2742005H, 2019ApJ...875..106C}. This problem can be resolved if the delay times of NSMs are sufficiently shorter than those of SNe Ia \citep{2021MNRASW}. 
In this study, we assume that 0.2 per cent of stars with 8--20 $M_{\odot}$ lead to NSMs. This rate corresponds to $\sim$10 {events} Myr$^{-1}$ in MW-equivalent galaxies. We also assume that each NSM produces 1.0 $\times$ $10^{-4}$ $M_{\odot}$ of Eu to reproduce the observed peak of [Eu/Fe] distribution ([Eu/Fe]\,$\sim+$0.4, \citealt{2020ApJS..249...30H}). This value corresponds to the total ejecta mass of 0.05 M$_{\sun}$ of one NSM, which is consistent with the values estimated from observations of GW170817 \citep[e.g.][]{2017PhRvL.119p1101A, 2017Natur.551...64A,2017ApJ...848L..18N,  2017PASJ...69..102T}.

Supernova-feedback models are implemented following \citet{2018MNRAS.477.1578H}. Thus, we transfer the terminal momentum of supernova remnants to surrounding gas particles when a supernova {explodes}. We assume that CCSNe and SNe Ia explode with 10$^{51}$ ergs. We also assume that 5 per cent of stars with 20--40 M$_{\sun}$ explode as broad-line type Ic supernovae (hypernovae) with energies of 10$^{52}$ erg. {This model does not include the energy feedback from stellar winds.} The energy of NSMs is not taken into account, as in \citet{2015ApJ...807..115S, 2015MNRAS.447..140V}, because the rate is significantly lower than CCSNe, and the estimation of energy from NSMs is highly uncertain \citep[e.g.][]{2013PhRvD..87b4001H}. \citet{2017MNRAS.471.2088S} confirmed that the energy of NSMs does not largely affect the abundances of Eu in UFDs. 

In this study, we implement the turbulence-induced metal-diffusion model \citep{2010MNRAS.407.1581S,2017ApJ...838L..23H,2017AJ....153...85S}. The $i$th elements ($Z_i$) injected into the interstellar medium (ISM) are diffused to the surrounding gas particles following the equation below: 
\begin{equation}
    \frac{{\rm{d}}Z_i}{{\rm{d}}t} = \nabla(D\nabla{Z_i}),~~ D = C_{\rm{d}}|S_{ij}|h^2,
\end{equation}
where $h$ is the smoothing length of the SPH particle, $S_{ij}$ is the trace-free shear tensor, and $C_{\rm{d}}$ is a scale factor for metal diffusion. In this study, we adopt $C_{\rm{d}}$ = 0.01. This value is calibrated using the observed $r$-process abundances in dwarf galaxies \citep{2017ApJ...838L..23H}. {Their study and \citet{2018ApJ...855...63H} have shown that models with $C_{\rm{d}}$ = 0.01 -- 0.1 can produce the scatter of [Ba/Fe] and [Zn/Fe] consistent with observations.}

\subsection{Initial conditions}

In this study, we performed a cosmological zoom-in simulation of an MW-like galaxy. A low-resolution pre-flight cosmological simulation of structure formation was performed with the $N$-body code \textsc{gadget-2} \citep{2005MNRAS.364.1105S}. The initial condition was generated by \textsc{music} \citep{2011MNRAS.415.2101H}. The box size is (36 $h^{-1}$ Mpc)$^3$. Density fluctuations on this scale are in the linear regime at $z$ = 0 \citep{2003MNRAS.345..429O}. We set the cosmological parameters as follows: $\Omega_{\rm{m}}\,=\,0.308$, $\Omega_{\Lambda}\,=\,0.692$, $\Omega_{\rm{b}}\,=\,0.0484$, $H_{0}\,=\,67.8$ km$\>$s$^{-1}$$\>$Mpc$^{-1}$, $\sigma_{8}\,=\,0.815$, $n_{\rm{s}}\,=\,0.968$ \citep{2016A&A...594A..13P}. 

We have selected the haloes from this simulation that can form MW-like galaxies. The selection criteria is as follows: (1) the total mass is from 5\,$\times$\,10$^{11}\,$M$_{\sun}$ to 2\,$\times$\,10$^{12}\,$M$_{\sun}$, (2) over half of the halo's total mass is developed by $z$ = 2, (3) spin parameters of the haloes are from 0.02 to 0.07 and (4) there are no haloes larger than 10$^{13}\,$M$_{\sun}$ within 3 $h^{-1}$\,Mpc from the simulated halo \citep[e.g.][]{2013ApJ...767..146I, 2016ApJ...818...10G}. We use the AMIGA halo finder (\textsc{ahf}) to find haloes and construct merger trees \citep{2004MNRAS.351..399G, 2009ApJS..182..608K}. We select high-resolution particles following the convex-hull structure to minimise the computational cost. High-resolution particles are collected within three times the virial radius of the main halo at $z$ = 0. The total number of high-resolution dark matter and gas particles is $8.0\times10^7$. High-resolution zoom-in hydrodynamics simulations were performed with \textsc{asura} (see Section \ref{sec:Code} for details). We have confirmed that the contamination from low-resolution particles is less than 1 per cent within the virial radius of the main halo \citep{2014MNRAS.437.1894O}. Table \ref{model} lists the parameters of the model halo adopted in this study.

\begin{table*}
	\centering
	\caption{Parameters of the selected halo. From left to right, columns show the name of the model, the total number of particles ($N$), the virial mass of the main halo ($M_{\rm{vir}}$), the virial radius ($R_{\rm{vir}}$) at $z$ = 0, the mass of one dark matter particle ($m_{\rm{DM}}$), the initial mass of one gas particle ($m_{\rm{gas}}$), the initial mass of one star particle {($m_{\rm{*}}$)} formed from gas particles with a mass of 1.3 $\times$ 10$^{4}$\,M$_{\sun}$, and the gravitational softening length for dark matter ($\epsilon_{\rm{DM}}$) and gas particles ($\epsilon_{\rm{g}}$).\label{model}}
	\label{tab:model}
	\begin{tabular}{cccccccrr} 
		\hline
		Model &
		$N$&
		$M_{\rm{vir}}$&$R_{\rm{vir}}$& $m_{\rm{DM}}$ & $m_{\rm{gas}}$ &
		 $m_{*}$ &
		 $\epsilon_{\rm{DM}}$&
		 $\epsilon_{\rm{g}}$\\
		 &&(M$_{\sun}$) &(kpc) &  (M$_{\sun}$) &  (M$_{\sun}$) &
		  (M$_{\sun}$) &
		 (pc)&(pc)\\		 
		\hline
		220lv12 &
	    9.2 $\times$ 10$^{7}$ &
		1.2 $\times$ 10$^{12}$ &
	    1.5 $\times$ 10$^{2}$  & 
		7.2 $\times$ 10$^{4}$  & 
	    1.3 $\times$ 10$^{4}$ &
	    4.5 $\times$ 10$^{3}$ &
		85 & 82\\
		\hline
	\end{tabular}
\end{table*}

\subsection{Definition of solar neighbourhood stars in the simulation}\label{sec:solar}

We define the position of the `sun' in the simulation to select star particles in a region comparable to the observations. Following the Sun's location and velocity in the Galaxy \citep{2016ARA&A..54..529B}, we set the `sun' in the simulation to the centre of the mass of stars located in (1) 5 to 11 kpc from the Galactic Centre, (2) $-$50 to 50 pc from the disc plane and (3) 150 to 350 km$\>$s$^{-1}$ of the azimuthal velocity. Following these criteria, the position of the `sun' in model 220lv12 is ($x,\,y,\,z$) = (4.4 kpc, $-$5.0 kpc, 0.0 kpc). This study mainly analyses stars within 5 kpc from this position to compare with observations. We also ignore stars within 3 kpc from the Galactic Centre to remove bulge stars, which are not typically targeted in observational surveys. Here we define star particles satisfying these conditions as solar neighbourhood stars. Although we do not explicitly categorize stars in this study, this region contains stars belonging to the thin, thick discs and stellar halo.

We note that all of the above assumptions for our simulations were made {\it in advance} of any comparison with the observed stellar kinematic and abundance information described below.  That is, no `tweaking' of the assumptions has been made, although this might be explored in the future.

\section{Results}\label{sec:results}

\subsection{Formation of a Milky Way-like galaxy}

\begin{figure}
 \centering
    \includegraphics[width=\columnwidth, bb = 0 0 450 350]{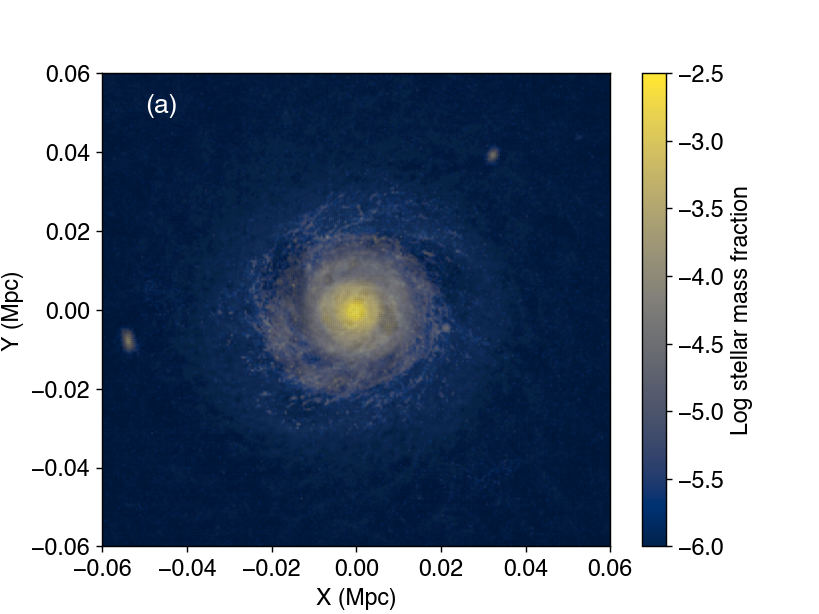}
  \includegraphics[width=\columnwidth, bb = 0 0 450 350]{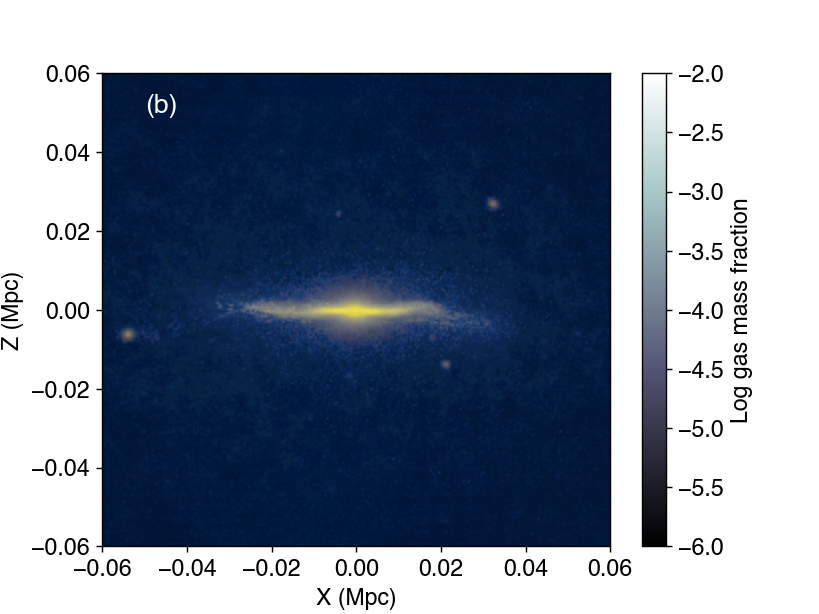}
\caption{The (a) face-on and (b) edge-on view of the stellar and gas distribution in the simulated MW-like galaxy at $z$ = 0. The colour scale represents the log-scale mass fraction of stars and gas in each grid {of 0.47 $\times$ 0.47 kpc$^2$}.}\label{fig:GasStars}
\end{figure}

The simulated galaxy has similar properties to the MW. Fig. \ref{fig:GasStars} shows face-on (Fig. \ref{fig:GasStars} (a)) and edge-on  (Fig. \ref{fig:GasStars} (b)) views of the stars and gas in 220lv12, indicating that a disc galaxy is formed in this simulation. The stellar and halo masses within the virial radius at $z\,=\,0$ are $8.44\,\times\,10^{9}\,M_{\odot}$ and $1.20\,\times\,10^{12}\,$M$_{\sun}$, respectively. The stellar mass-halo mass ratio at the time of the peak stellar mass is 0.007. This value is similar to the stellar mass-halo mass ratios of the MW \citep{2016ARA&A..54..529B} and other cosmological simulations of MW-like galaxies \citep[e.g.][]{2011ApJ...742...76G, 2017MNRAS.467..179G, 2018MNRAS.480..800H, 2020MNRAS.498.1765F, 2021ApJ...906...96A, 2021MNRAS.503.5826A}. Although the final stellar mass is smaller than the MW ($\approx$ 5\,$\pm\,1$\,$\times$\,10$^{10}\,$M$_{\sun}$), {we minimize the effects of the difference in the stellar mass by adjusting the position of the `sun' to avoid selecting outer-disc stars, which are not presently searched for RPE stars. In Section \ref{sec:comparison}, we compare our results to previous simulations of MW-like galaxies.} 
Below we compare the results from this simulation to the observations of pertinent elemental-abundance ratios and the stellar kinematics of RPE stars.

\begin{figure*}
  \centering
  \includegraphics[width=\columnwidth, bb = 0 0 450 350]{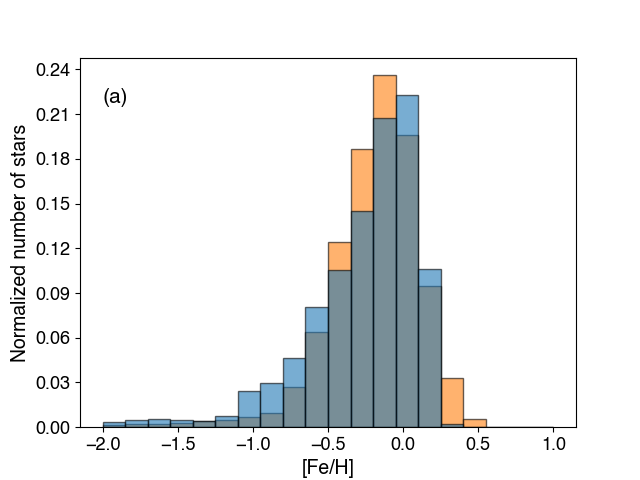}
    \includegraphics[width=\columnwidth, bb = 0 0 450 350]{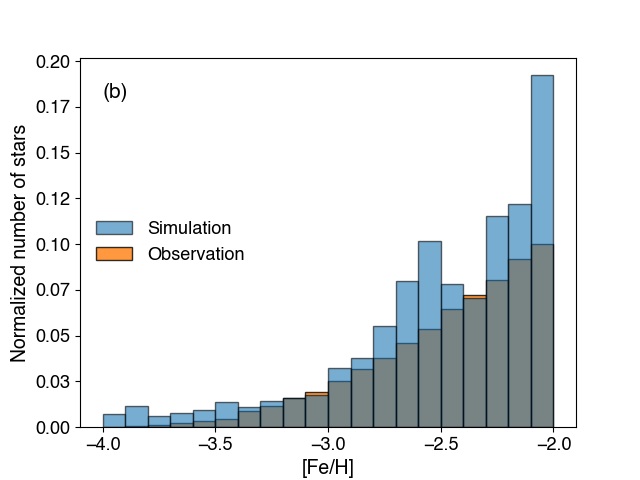}
\caption{Metallicity distribution functions (MDFs) of {(a) the MW's solar neighbourhood stars and (b) halo stars. Blue and orange histograms represent simulated and observed stars, respectively.} The simulated data are the MDF of star particles with distances within 5 kpc from the `sun' and outside of 3 kpc from the Galactic Centre in the simulation. Observed data {in panel (a)} are taken from GALAH DR3 \citep{2021MNRAS.506..150B} with
    $\texttt{flag\_sp}=0$,
    $\texttt{snr\_c3\_iraf}>50$,
    $\texttt{flag\_fe\_h}=0$. {Observed data in panel (b) are taken from the Galactic halo stars of \citet{2020MNRAS.492.4986Y}.} See text for details.}\label{fig:MDF}
\end{figure*}

The metallicity distribution function (MDF) of star particles within 5 kpc from the `sun' and outside of 3 kpc from the Galactic Centre in our simulation shown in Fig. \ref{fig:MDF} is similar to that of the MW. {Fig. \ref{fig:MDF} (a) compares simulated and observed MDF taken from GALAH DR3 for [Fe/H]$\,>\,-2$ \citep{2021MNRAS.506..150B}. Since GALAH data do not obtain accurate metallicity estimates for stars with [Fe/H]$\,<-$2, we compare the MDFs for stars with [Fe/H]$\,>-2$.} The simulated MDF of solar neighbourhood stars {for [Fe/H]$\,>\,-2$} exhibits a median [Fe/H] $= -$0.17, with an interquartile range of 0.44 dex. By comparison, the MDF based on GALAH DR3 exhibits a median [Fe/H] = $-0.16$ and interquartile range of 0.35 dex \citep{2021MNRAS.506..150B}. 

{The lower metallicity end of simulated MDFs is also similar to observations. Fig. \ref{fig:MDF} (b) compares simulated and observed halo star MDFs \citep{2020MNRAS.492.4986Y} for stars with [Fe/H]$\,<\,-2$. Although a two-sample Kolmogorov–Smirnov (KS) test for the observed and simulated MDFs rejects the null hypothesis that the two distributions are drawn from the same parent populations at a significance level of 0.05 ($p = 0.04)$, both data exhibit decreasing trends towards lower metallicity. Peaks in the simulated MDFs at [Fe/H] = $-2.0$ and $-2.6$ are possibly due to the contribution of the relatively massive accreted components (e.g. haloes 3, 4, and 5 in Table \ref{tab:group}).}

\begin{figure}
 \includegraphics[width=\columnwidth, bb = 0 0 450 350]{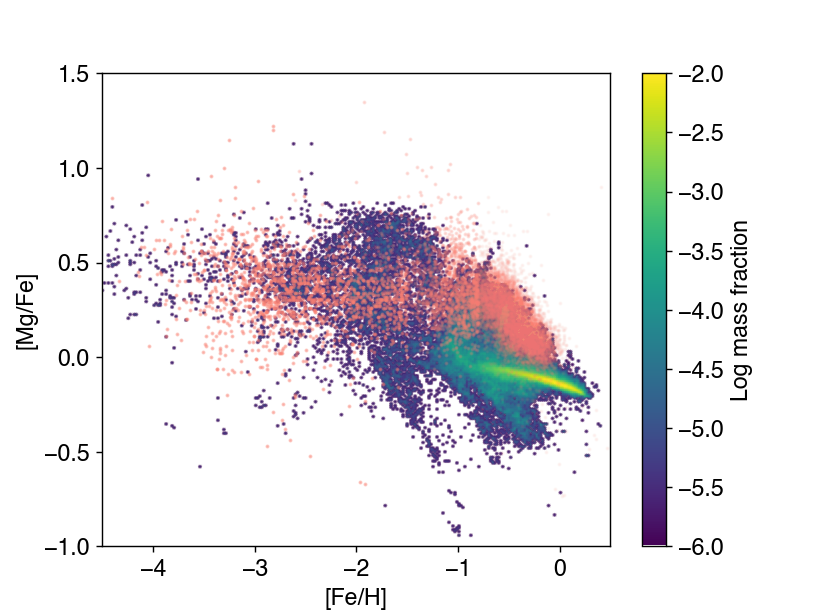}
    \caption{[Mg/Fe] versus [Fe/H] at $z$ = 0 in the simulation. Stars in the simulation are within 5 kpc distance from the `sun' and outside of 3 kpc from the Galactic Centre. The colour scale represents the log-scale mass fraction of stars in each grid {of 0.02 $\times$ 0.01 dex$^2$}, from 10$^{-6}$ (purple) to 10$^{-4}$ (yellow). The orange represent observed abundances in the MW taken from the SAGA database \citep{2008PASJ...60.1159S, 2011MNRAS.412..843S, 2017PASJ...69...76S, 2013MNRAS.436.1362Y} for [Fe/H]$\,<-1.5$ and GALAH DR3 \citep{2021MNRAS.506..150B} with
    $\texttt{flag\_sp}=0$,
    $\texttt{snr\_c3\_iraf}>50$,
    $\log{g}<1.9$,
    $\texttt{flag\_fe\_h}=0$, and 
    $\texttt{flag\_Mg\_fe}=0$.  Note that RPE stars from \citet{shank2022c} are not included in this figure.}\label{fig:MgFe}
\end{figure}

Our simulation also exhibits general trends of [Mg/Fe] ratios that are similar to the observations. Fig. \ref{fig:MgFe} shows the [Mg/Fe] ratios, as a function of [Fe/H], for solar neighbourhood stars simulated in 220lv12. For [Fe/H] $<-2$, the [Mg/Fe] ratios are constant ($\sim+$0.5) with a small scatter ($\sigma\,=\,0.19$ dex). Most of these stars are associated with the stellar halo. As is expected,  the iron production from SNe Ia decreases the [Mg/Fe] ratios for [Fe/H] $>-2$. Most stars with [Mg/Fe]$\,<\,$0 and $-2\,<\,$[Fe/H]$\,<-1$ come from the accreted component, halo 5, discussed in Section \ref{sec:discussion}. Note that our assumption of the minimum timescale for SNe Ia (40 Myr) would shift the metallicity of the `knee' in [Mg/Fe]. This assumption {is likely responsible for the constant [Eu/Fe] ratios at high metallicity due to small differences in the delay time distribution of SNe Ia and NSMs. This issue is discussed further in Section \ref{sec:caveats}}.

\subsection{The enrichment of Eu}\label{sec:Eu}

\begin{figure}
	 \includegraphics[width=\columnwidth, bb = 0 0 420 330]{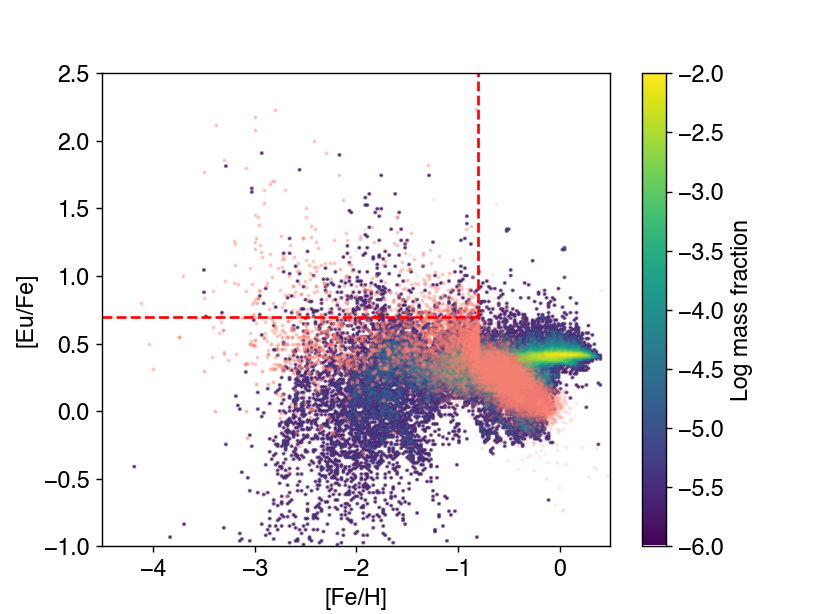}
    \caption{[Eu/Fe] as functions of [Fe/H] at $z$ = 0. Stars in the simulation are taken within 5 kpc from the `sun' and outside of 3 kpc from the Galactic Centre. The colour scale represents the log-scale mass fraction of stars in each grid {of 0.02 $\times$ 0.01 dex$^2$} from 10$^{-6}$ (purple) to 10$^{-2}$ (yellow). Orange plots represent observed abundances for stars in the MW taken from RPA \citep{2020ApJS..249...30H}, GALAH DR3 \citep{2021MNRAS.506..150B} with $\texttt{flag\_sp}=0$,
    $\texttt{snr\_c3\_iraf}>50$,
    $\log{g}<1.9$,
    $\texttt{flag\_fe\_h}=0$, 
    $\texttt{flag\_Eu\_fe}=0$ and \citet{shank2022c} for stars with [Fe/H] $< -0.8$ and [Eu/Fe] $> +0.3$. {\citet{shank2022c} are composed of the literature data taken from \href{https://jinabase.pythonanywhere.com/}{JINAbase} \citep{2018ApJS..238...36A} and GALAH DR3. The upper left region indicated by the red dashed line is plotted in Fig. \ref{fig:EuFeRPE}.  }} \label{fig:EuFe}
\end{figure}

Fig. \ref{fig:EuFe} shows [Eu/Fe], as a function of [Fe/H], for solar neighbourhood star particles. As shown in this figure, there exists a star-to-star scatter in the [Eu/Fe] ratios of over 2 dex, similar to the observations for [Fe/H] $\lesssim -2$. Here we define stars with [Eu/Fe] $>$ +0.7 as $r$-II stars. The fraction of $r$-II stars is 8.7 per cent in this simulation, which is similar to the fraction reported in the RPA \citep[8.3 per cent,][]{2020ApJS..249...30H} -- note that both of these values are computed for stars with [Fe/H]$\,<-2.5$ and [Eu/Fe]$\,>-0.8$. {SNe Ia dominantly start to contribute for [Fe/H]$\,>\,-2$ in this simulation.} Owing to the delay times adopted for NSMs, the RPE stars in the simulation are formed with [Fe/H] $\gtrsim$ $-$3.5. 

\begin{figure}
  \centering
  \includegraphics[width=\columnwidth, bb = 0 0 450 350]{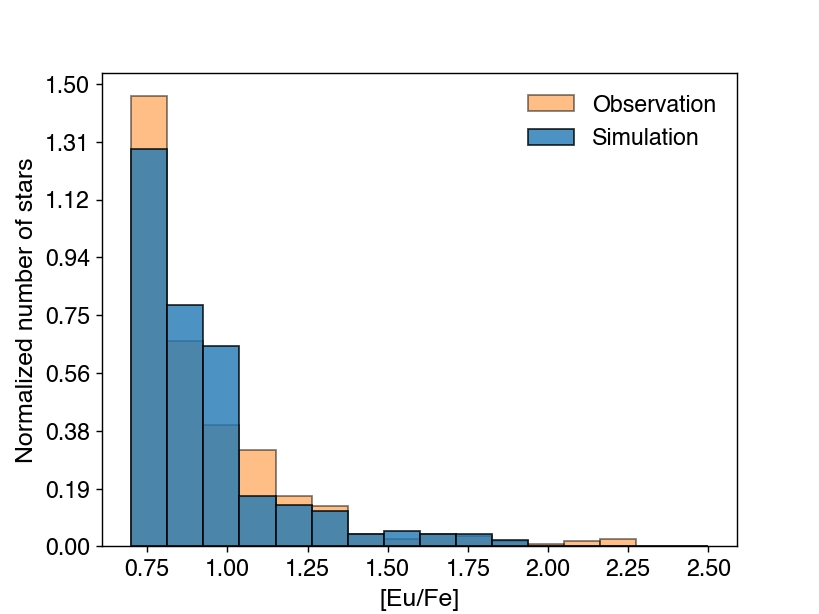}
\caption{The [Eu/Fe] distribution of the observed (orange) and simulated (blue) $r$-II stars with [Fe/H]$\,<-0.8$. The simulated data are the solar neighbourhood stars defined in Section \ref{sec:solar}. Observed data are taken from \citet{shank2022c} and GALAH DR3 \citep{2021MNRAS.506..150B}.}
\label{fig:EuFeRPE}

\end{figure}

The simulated distribution of [Eu/Fe] for metal-poor $r$-II stars provides a good match to that of the observations {for $r$-II stars}, as shown in Fig. \ref{fig:EuFeRPE} for $r$-II stars with {[Fe/H]$\,< -0.8$}. The metallicity cut ({[Fe/H]$\,< -0.8$}) is based on \citet{shank2022c} {to be the same as the data taken by the RPA \citep{2020ApJS..249...30H}}. As can be appreciated from inspection, the simulated [Eu/Fe] distribution of metal-poor $r$-II stars reproduces the decreasing number fraction of these stars towards higher [Eu/Fe] ratios. The skewness of the [Eu/Fe] distribution for the simulation and the observations are 1.9 and 2.2, respectively, indicating that the two distributions are quite similar. From the application of a two-sample KS test for the observed and simulated [Eu/Fe] distributions of $r$-II stars, we are unable to reject the null hypothesis that the two distributions are drawn from the same parent populations ($p = 0.32)$.

\subsection{Kinematics of $r$-process-enhanced stars}

\begin{figure}
	 \includegraphics[width=\columnwidth, bb = 0 0 450 350]{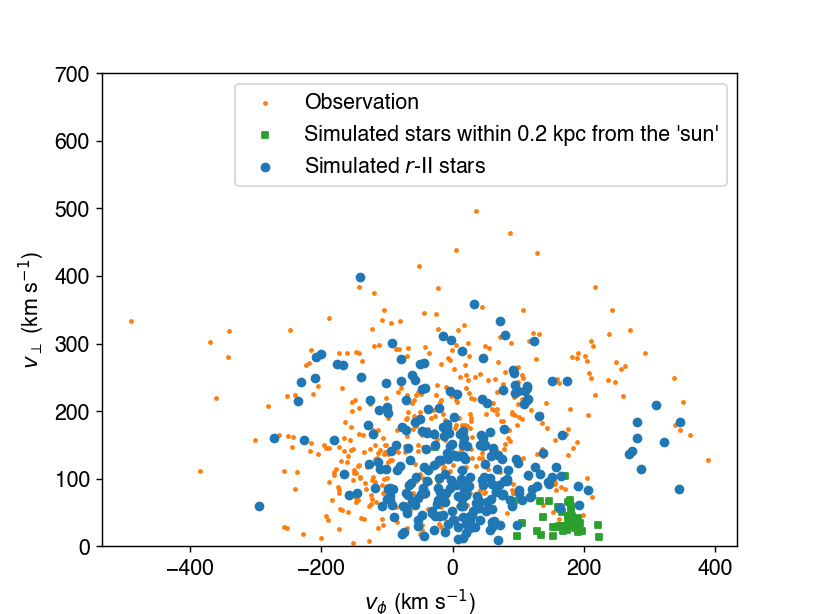}
    \caption{Toomre diagram for the observations (orange filled circles), simulated stars with a distance of 0.2 kpc from the `sun' (green squares), and simulated $r$-II stars with [Fe/H] $<-1$ (blue filled circles). The observed data are taken from \citet{shank2022c}. The velocities are plotted with a cylindrical coordinate system ($V_{R}$, $V_{\phi}$, $V_{z}$), where $V_{\phi}$ is the azimuthal velocity and $V_{\perp}$ is defined as {$(V_{R}+V_{z})^{1/2}$}.}
    \label{fig:Toomre}
\end{figure}

Few $r$-II stars with [Fe/H] $<-1$ exhibit disc-like orbits. Fig. \ref{fig:Toomre} is the Toomre diagram for stars around the `sun' in our simulation. In the figure, we plot the solar neighbourhood disc and $r$-II stars with [Fe/H] $<-1$ within 0.2 kpc and 5 kpc from the `sun' in the simulation, respectively, to compare with the observations \citep{shank2022c}. In the simulation, 92 per cent of stars lie beyond a 100 km$\>$s$^{-1}$ velocity difference from the `sun'. This result indicates that most RPE stars are kinematically associated with the stellar halo system. The fraction of simulated $r$-II stars with prograde and retrograde orbits are 55 per cent and 45 per cent, respectively,  similar to the observations. \citet{shank2022c} find that the fractions of prograde and retrograde $r$-II stars are 65 per cent and 35 per cent, respectively. They also find that only 8 per cent of $r$-II stars exhibit disc-like orbits.

For [Fe/H] $>-1$, all of the simulated $r$-II stars are formed \textit{in situ}. In this metallicity range, 62 per cent of the simulated $r$-II stars lie within a 100 km$\>$s$^{-1}$ velocity difference from the ‘sun’, i.e. there are a larger fraction of stars with disc-like kinematics compared to stars with [Fe/H] $<-1$. Moreover, the fraction of stars with prograde orbits is larger than for the lower metallicity stars. For [Fe/H] $>-1$, 84 per cent of the stars exhibit prograde orbits. If this were the case, a greater fraction of $r$-II stars with prograde orbits would be observed at high metallicity. As we discuss in Section \ref{sec:caveats}, the fraction of $r$-II stars could be over-estimated in this simulation due to the assumed level of {metal mixing} and the delay-time distribution of NSMs. Previous observations suggest that relatively few $r$-II stars are found in this metallicity range. Therefore, we do not include these stars in Fig. \ref{fig:Toomre}.
 
\section{Discussion}\label{sec:discussion}

Here we discuss the formation of $r$-II stars and their chemo-dynamical properties. 

Table \ref{tab:group} shows the list of the haloes that formed $r$-II stars and were accreted into the main halo (halo 0) in our simulation. These are ordered by the snapshot in time when the first $r$-II star is detected in each.

\begin{table*}
	\centering
	\caption{List of haloes where $r$-II stars are formed. From left to right, the columns show the halo number, the time when the halo formed ($t_{\rm{form}}$), the time when the first $r$-II star formed in the halo ($t_{\rm{rII}}$), the time when the halo is accreted into the main halo ($t_{\rm{acc}}$), and the total mass ($M_{\rm{tot}}$), gas mass ($M_{\rm{gas}}$), and stellar mass ($M_{\rm{*}}$) at the time of the first $r$-II star formation in a given halo. The time listed in this table is the one from the beginning of the simulation. Halo 0 is the main halo (\textit{in situ} component) of this simulation. {The halo number is assigned based on the order of the snapshots in which the first $r$-II star is formed.}}
	\label{tab:group}
	\begin{tabular}{ccccrrr} 
		\hline
		Halo No.&
		$t_{\rm{form}}$&
		$t_{\rm{rII}}$&
		$t_{\rm{acc}}$&
		$M_{\rm{tot}}$&
		$M_{\rm{gas}}$&
        $M_{\rm{*}}$
		\\
		 &(Gyr) & (Gyr) &  (Gyr) &  ($10^9\,$M$_{\sun}$) &
		  ($10^8\,$M$_{\sun}$) &
		 ($10^6\,$M$_{\sun}$)\\		 
		\hline
0  & 0.24 & 0.43 & - & 5.02 & 3.60 & 7.34 \\
1  & 0.17 & 0.40 & 0.73 & 0.36 & 0.47 & 0.05 \\
2  & 0.18 & 0.39 & 0.49 & 0.38 & 0.36 & 0.12 \\
3 & 0.18  & 0.44 & 0.76 & 6.49 & 4.66 & 4.28 \\
4 & 0.15 & 0.45 & 1.27 & 2.91 & 1.88 & 2.26 \\
5 & 0.13  & 0.53 & 1.12 & 3.65 & 3.31 & 0.96\\
6 & 0.36 & 0.53 & 0.57 & 1.29 & 1.14 & 0.37\\
7 & 0.15 & 0.55 & 0.68 & 0.30 & 0.21 & 0.24\\
8 & 0.13 & 0.58 & 2.46 & 2.55 & 2.87 & 4.03\\
9 & 0.18 & 0.58 & 8.44 & 2.85 & 1.68 & 1.69\\
10 & 0.13 & 0.59 & 2.48 & 2.53 & 2.85 & 1.64\\
11 & 0.39 & 0.59 & 0.94 & 1.04 & 2.11 & 0.45\\
12 & 0.44 & 0.71 & 1.69 & 3.57 & 0.68 & 0.50\\
13 & 0.39 & 0.93 & 3.43 & 3.08 & 1.92 & 6.12\\
14 & 0.47 & 1.07 & 4.57 & 4.24 & 6.49 & 1.75\\
15 & 0.33 & 1.23 & 1.64 & 1.37 & 0.73 & 0.50\\
16 & 0.38 & 1.67 & 9.78 & 33.99 & 20.27 & 14.38\\
		\hline
	\end{tabular}
\end{table*}

\subsection{The formation of $r$-II stars}\label{sec:form}

\begin{figure}
	 \includegraphics[width=\columnwidth, bb = 0 0 410 320]{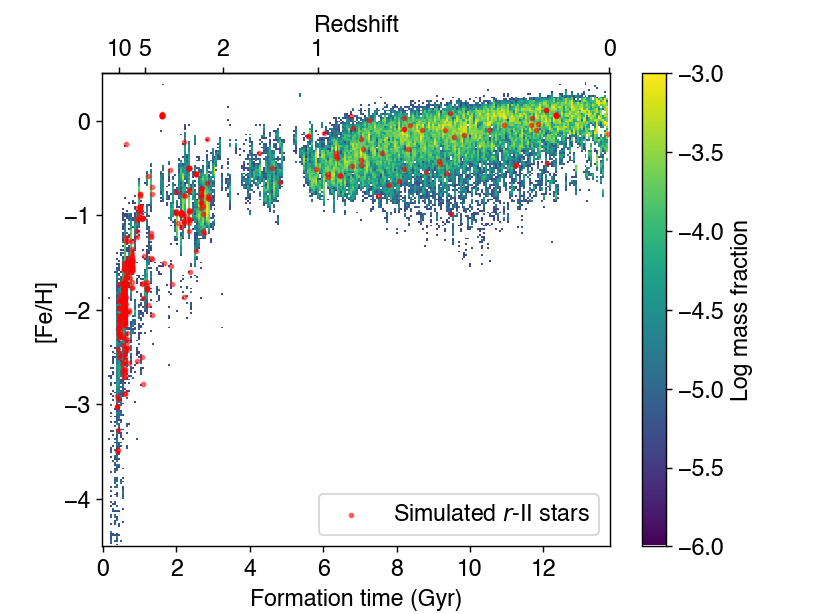}
    \caption{[Fe/H], as a function of the formation time, of the solar neighbourhood simulated stars. The colour scale represents the log-scale mass fraction of stars in each grid {of 0.05 Gyr $\times$ 0.02 dex}, from 10$^{-6}$ (purple) to 10$^{-3}$ (yellow). The red dots denote the $r$-II stars formed in the simulation.}
    \label{fig:FeHTime}
\end{figure}

In our simulation, most of the $r$-II stars are formed in the early epochs of galaxy evolution. Fig. \ref{fig:FeHTime} shows [Fe/H] as a function of the time from the beginning of the simulation. The discontinuous distribution of stars seen around 2 to 6 Gyr is due to the suppression of star formation by supernova feedback. {Star formation in galaxies is also affected by a cold accretion flow \citep[e.g.][]{2005MNRAS.363....2K}.} As seen in this figure, over 90 per cent of the $r$-II stars are formed within 4 Gyr from the beginning of the simulation ($z\,>\,1.6$). All of the stars with [Fe/H] $<-$1.5 are formed at $z\,>\,1.6$. This epoch corresponds to the time when the stellar halo is formed, suggesting that most of these stars come from {dwarf galaxies that are later accreted to the main halo}. Since the Galactic disc is formed in the later stage of the galaxy evolution, this result explains the halo-like kinematics seen in Fig. \ref{fig:Toomre}. 

The $r$-II stars with [Fe/H]$\,\geq-1.5$ are formed throughout the entire history of galaxy evolution. Most of these stars are formed in the main progenitor halo or in the disc, where they were locally enhanced in $r$-process elements. Since NSMs eject prolific amounts of $r$-process elements, some RPE stars are still formed in the later phases of galaxy evolution. 

\begin{figure*}
	 \includegraphics[width=\columnwidth, bb = 0 0 450 400]{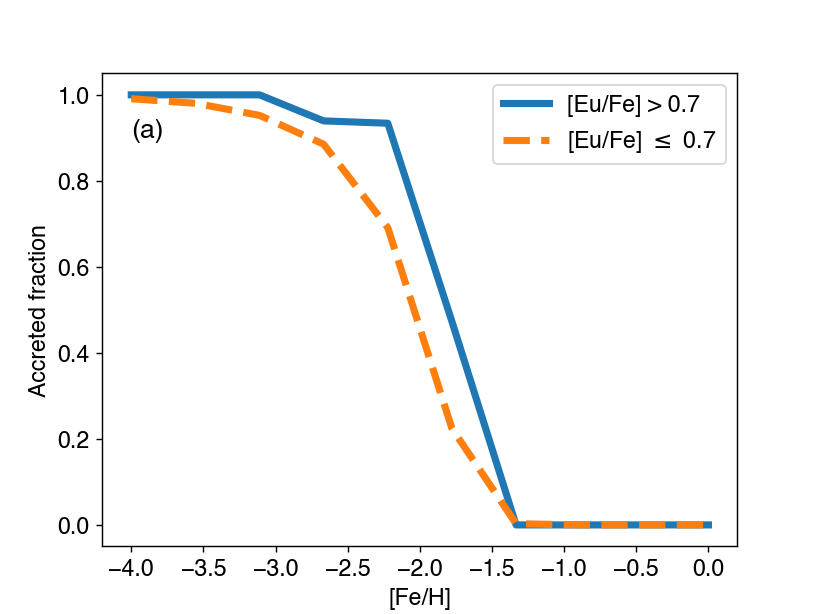}
	\includegraphics[width=\columnwidth, bb = 0 0 450 300]{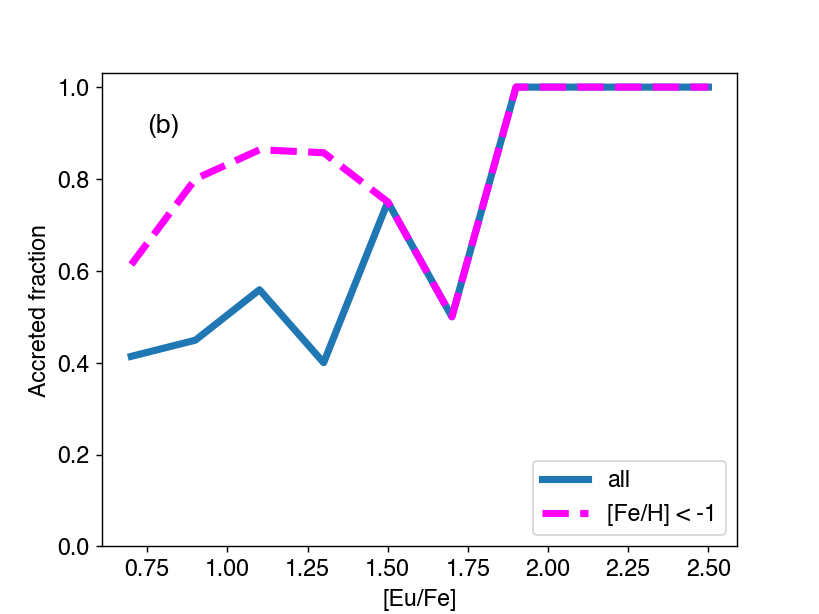}
    \caption{The number fraction of stars from accreted components, as a function of (a) [Fe/H] (blue solid curve: stars with [Eu/Fe] $>$ +0.7; orange dashed curve: stars with [Eu/Fe] $\leq$ +0.7) and (b) [Eu/Fe] (blue solid curve: all $r$-II stars, magenta dashed curve: $r$-II stars with [Fe/H] $<-1$).}
    \label{fig:f_acc}
\end{figure*}

The low-metallicity $r$-II stars tend to be formed in accreted components, as shown in Fig. \ref{fig:f_acc} (a). The number fraction of stars formed in accreted components, as a function of [Fe/H], demonstrates that there is an increasing trend of accreted fractions towards lower metallicity. Over 80 per cent of the $r$-II stars with [Fe/H] $<-2$ are formed in accreted components, while all $r$-II stars with [Fe/H] $>-1$ are formed \textit{in situ}. 

The accreted fraction of $r$-II stars is larger than that of non-RPE stars in the metallicity range $-2.5\,<\,$[Fe/H]$\,<-1.5$. At [Fe/H] = $-$2.2, 96 per cent (71 per cent) of $r$-II (non-RPE) stars are formed in the accreted components. This result is consistent with \citet{2020MNRAS.494.4867V}, who found that 91 per cent of $r$-II stars with [Fe/H] $<-$2 were formed in accreted components, using data from the Auriga simulation \citep{2017MNRAS.467..179G}. This fraction was larger than for all very metal-poor stars (78 per cent).

Fig. \ref{fig:f_acc} (b) shows the accreted fractions as a function of [Eu/Fe]. From inspection, there exists an increasing trend of the accreted fraction towards higher [Eu/Fe] ratios for all $r$-II stars. This trend is mainly due to the contribution of \textit{in situ} $r$-II stars with [Fe/H] $>-$1. These stars decrease the accreted fraction of $r$-II stars with [Fe/H] $<-$1.5. Excluding these stars increases the accreted fraction (the magenta dashed curve in Fig. \ref{fig:f_acc} (b)).

\begin{figure*}
	 \includegraphics[width=\columnwidth, bb = 0 0 450 400]{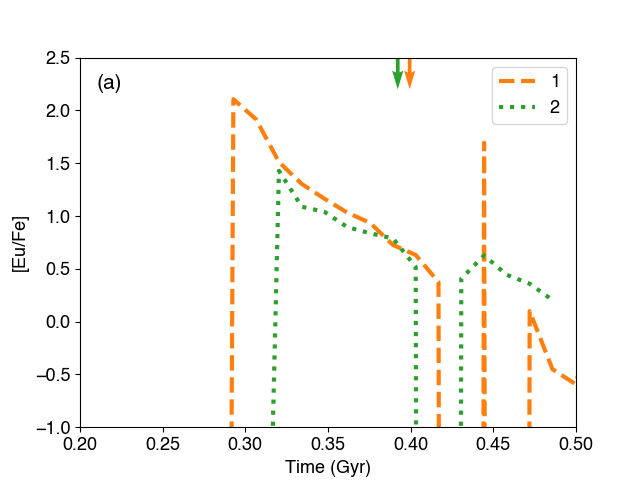}
	\includegraphics[width=\columnwidth, bb = 0 0 450 300]{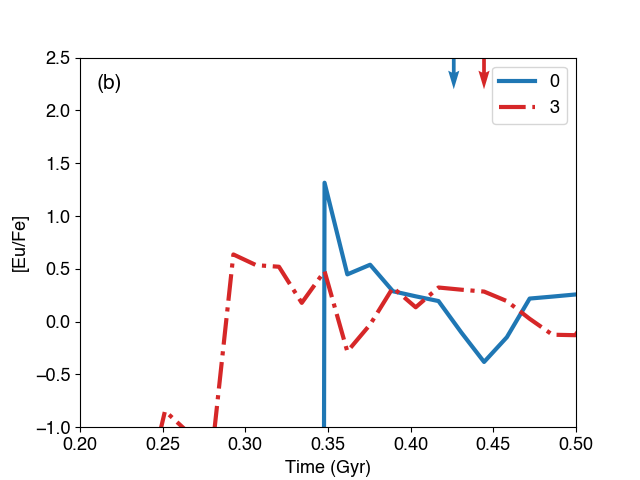}
    \caption{Time evolution of the mean gas-phase [Eu/Fe] ratios. Panel (a) represents halo 1 (orange dashed curve) and halo 2 (green-dotted curve), while panel (b) shows halo 0 (main progenitor halo, blue-solid curve) and halo 3 (red dash-dotted curve). Arrows indicate the formation time of $r$-II stars in each component (haloes 0: blue, 1: orange, 2: green, 3: red).}
    \label{fig:gas}
\end{figure*}

\begin{figure*}
	 \includegraphics[width=\columnwidth, bb = 0 0 450 400]{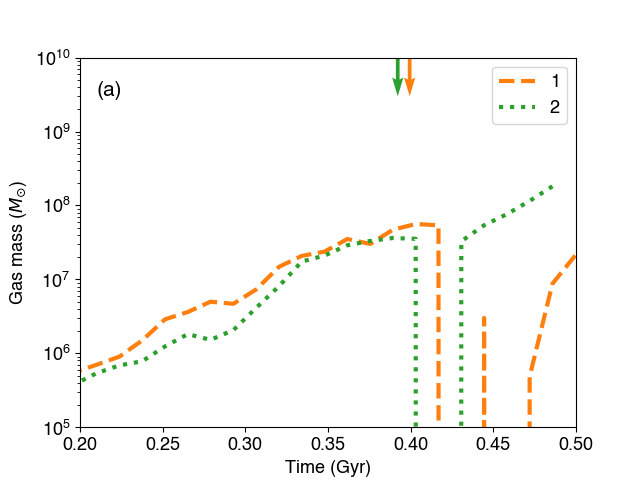}
	\includegraphics[width=\columnwidth, bb = 0 0 450 300]{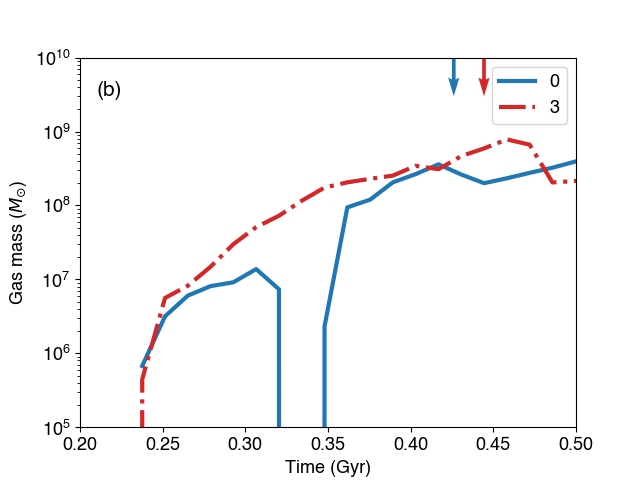}
    \caption{Similar to Fig. \ref{fig:gas} but for gas mass, as a function of time, in each halo.}
    \label{fig:gasmass}
\end{figure*}

The $r$-II stars are formed in gas clumps enhanced with $r$-process elements. Fig. \ref{fig:gas} (a) shows the time evolution of the average gas-phase [Eu/Fe]. The time variation of [Eu/Fe] depends on the gas mass of the building blocks. In the case of small building blocks (e.g. haloes 1 and 2), the mean [Eu/Fe] largely varies owing to the frequency of NSM events. The maximum mean [Eu/Fe] in halo 1 is +2.11 at the time from the beginning of the simulation ($t\,=\,0.29\,$Gyr). At the time of $r$-II star formation ($t$ = 0.40 Gyr), it is still [Eu/Fe] = +0.73. At $t$ = 0.29 Gyr, the gas mass of this galaxy is 4.7$\>\times\>10^6\>M_{\sun}$ (Fig.\ref{fig:gasmass} (a)). In such a small system, one NSM can enhance most of the gas contained in the halo; stars formed in this phase may have large enhancements of $r$-process elements as a result.

This mechanism is similar to $r$-II star formation in UFDs. Cosmological zoom-in simulations have shown that the small gas mass of UFDs is sufficient to enhance the $r$-process abundance after an NSM to form $r$-II stars \citep{2017MNRAS.471.2088S, 2020MNRAS.494..120T}, due to the lack of significant dilution by the gas reservoir. \citet{2021MNRAS.506.1850J} have shown that the most important parameter to form $r$-II stars is the timescale of $r$-II star formation after the Eu enrichment. Their results suggested that $r$-II stars need to be formed 10 to 100 Myr after an NSM event to keep the gas-phase Eu abundance ratio over [Eu/Fe] $>$ +1. The condition to form $r$-II stars in our simulation is similar. As shown in the orange dashed curve in Fig. \ref{fig:gas} (a), the average [Eu/Fe] drops within $\approx\,$100 Myr because of SN feedback, metal diffusion, and newly accreted gas. 

The $r$-II stars found in relatively large building blocks and the main progenitor halo are formed in gas clumps that are locally enhanced in $r$-process elements. The red dash-dotted curve in Fig. \ref{fig:gas} (b) shows the mean gas-phase [Eu/Fe] ratio in halo 3. The gas mass of this galaxy reaches $4.7\>\times\>10^{8}\>M_{\sun}$ at the time of the formation of $r$-II stars (Fig. \ref{fig:gasmass} (b)). At the time of the $r$-II star formation, the total mass of this halo is larger than the main progenitor halo (halo 0, Table \ref{tab:group}). Most of the mass of this halo is stripped when it accretes to halo 0. In this halo, the average gas-phase [Eu/Fe] is not highly enhanced when $r$-II stars are formed. This result suggests that an NSM cannot solely enhance $r$-process elements in haloes with a gas mass larger than $\sim$$10^{8}\>M_{\sun}$.

\begin{figure}
	 \includegraphics[width=\columnwidth, bb = 0 0 450 450]{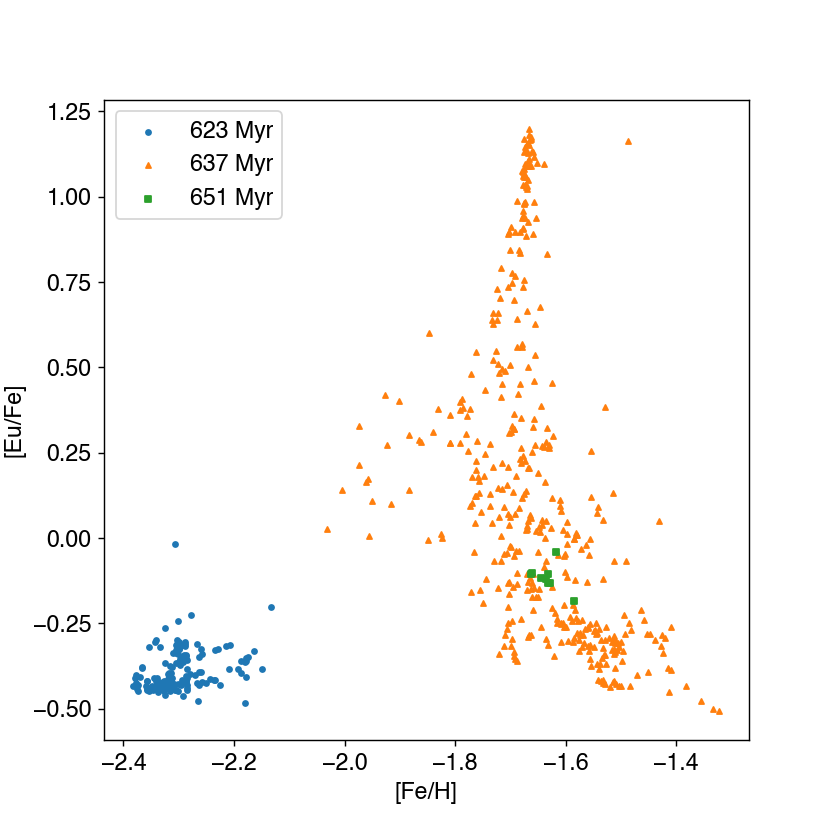}
    \caption{Evolution of gas-phase [Eu/Fe], as a function of [Fe/H], within 200 pc from the formation site of an \textit{in situ} $r$-II star. The blue circle, orange triangles and green squares represent gas particles in this region at $t$ = 623, 637 and 651 Myr, respectively.}
    \label{fig:EuFegas}
\end{figure}

Fig. \ref{fig:EuFegas} shows the gas-phase [Eu/Fe], as a function of [Fe/H], within 200 pc from a progenitor gas particle of \textit{in situ} $r$-II stars. We point to the most $r$-process enhanced \textit{in situ} star ([Fe/H] = $-$1.75, [Eu/Fe] = +1.76) from our sample for this analysis. Before star formation initiates, all gas particles have [Eu/Fe] $<$ 0 ($t$ = 623 Myr). After that, some gas particles show [Eu/Fe] $>$ +1 in this region owing to a nearby NSM ($t$ = 637 Myr); the [Fe/H] ratios are increased due to a nearby supernova. The number of gas particles is increased compared to the previous snapshot because the gas particles are in a converging flow that enters this region. After star formation initiates, several gas particles are consumed to form stars, and the [Eu/Fe] ratios are decreased due to metal diffusion.

One NSM can cause local enhancements of $r$-process elements in gas clumps. In this simulation, an NSM typically distributes $r$-process elements to gas clumps with gas mass $\sim$10$^6\,$M$_{\sun}$. Inside these clumps, there are variations of [Eu/Fe] ratios (Fig. \ref{fig:EuFegas}). Gas particles closer to the NSM are more enhanced in $r$-process elements than more distant particles. This coincidence is highly rare. Therefore, this mechanism of forming $r$-II stars is only possible in relatively massive building blocks that can host sufficient numbers of NSMs. This result implies that if stars are formed in a highly $r$-process-enhanced region near an NSM, high metallicity $r$-II stars could be formed.

The formation of $r$-II stars in relatively massive building blocks found in this study could be used to understand the presence of $r$-II stars found in massive satellite galaxies (e.g. LMC, SMC, Fornax). \citet{2021AJ....162..229R} found that six out of nine (two out of four) metal-poor LMC (SMC) giants are classified as $r$-II stars. In the Fornax dSph, three $r$-II stars with +1.25 $\leq$ [Eu/Fe] $\leq$ +1.45 have been confirmed \citep{2021ApJ...912..157R}. Although these stars have relatively high metallicity ($-1.3\,\leq$ [Fe/H] $\leq-0.8$), their chemistry clearly shows the typical $r$-process abundance pattern. These $r$-II stars could be formed in locally $r$-process-enhanced gas clumps.

\begin{figure}
	 \includegraphics[width=\columnwidth, bb = 0 0 450 400]{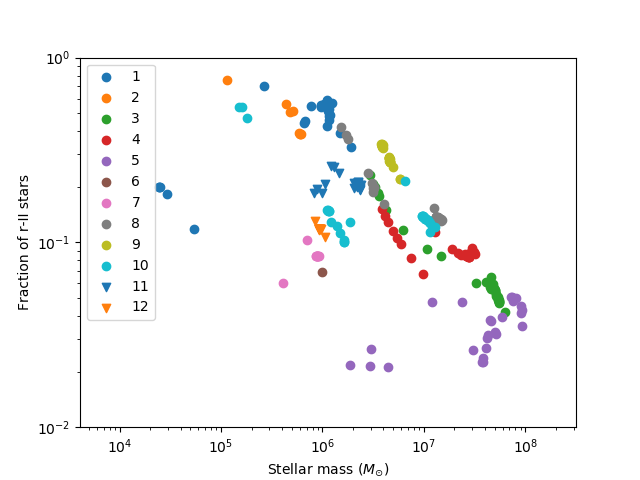}
    \caption{The number fraction of $r$-II stars, as a function of the stellar mass of building-block galaxies. The same colour and symbol indicate the building-block galaxy in different snapshots from 0.13 to 1.06 Gyr from the beginning of the simulation; the snapshots are separated by 13.8 Myr. The circles represent haloes from 1 to 10, and the inverted triangles represent haloes from 11 and 12. See legend for details.}
    \label{fig:f_rII}
\end{figure}

Lower mass building blocks tend to have larger fractions of $r$-II stars. Fig. \ref{fig:f_rII} shows the number fraction of $r$-II stars in building-block galaxies. As seen in this figure, there are clear increasing trends of this fraction towards the lower stellar mass. Since lower mass galaxies tend to have lower gas mass, an NSM enhances the abundances of $r$-process elements significantly due to the lack of dilution (Fig. \ref{fig:gas}). Subsequent star formation from $r$-process-depleted gas decreases the fraction of $r$-II stars in each halo. The scatter is mainly caused by a newly contributing NSM in each halo. The $r$-II stars formed around the ejecta of the NSM enhance the fraction of $r$-II stars and stellar mass of the building block galaxy.

The above results could explain the observed trend of $r$-process enhancement in dwarf galaxies. In the Reticulum II UFD, seven out of nine stars observed in this galaxy are enhanced in $r$-process elements \citep{2016Natur.531..610J}. In contrast, classical dSphs tend to have smaller fractions of RPE stars than in Reticulum II \citep[e.g.][]{2015ARA&A..53..631F}. There are 41 out of 165 $r$-II stars in Fornax \citep{2010A&A...523A..17L}, Ursa Minor \citep{2001ApJ...548..592S, 2004PASJ...56.1041S}, Carina \citep{2003AJ....125..684S, 2017ApJS..230...28N}, Draco \citep{2009ApJ...701.1053C}, and Sculptor \citep{2003AJ....125..684S, 2005AJ....129.1428G} dSphs \citep[counted from the SAGA database,][]{2008PASJ...60.1159S, 2011MNRAS.412..843S, 2017PASJ...69...76S, 2013MNRAS.436.1362Y}. These results imply that $r$-II stars in classical dSphs are formed by local inhomogeneities, while those in UFDs are formed in a halo entirely enhanced in $r$-process elements. {Another possibility is that $r$-II stars formed in UFD-mass galaxies that are accreted by other dwarf galaxies. This could happen in relatively large dwarf galaxies, such as LMC and Fornax \citep[e.g.][]{2021ApJ...912..157R}. Detailed studies of the spatial distribution and kinematics of $r$-II stars in surviving dwarf galaxies could help resolve this issue.}

\subsection{Chemo-dynamical properties of $r$-II stars}

\begin{figure*}
	 \includegraphics[width=\columnwidth, bb = 0 550 550 0]{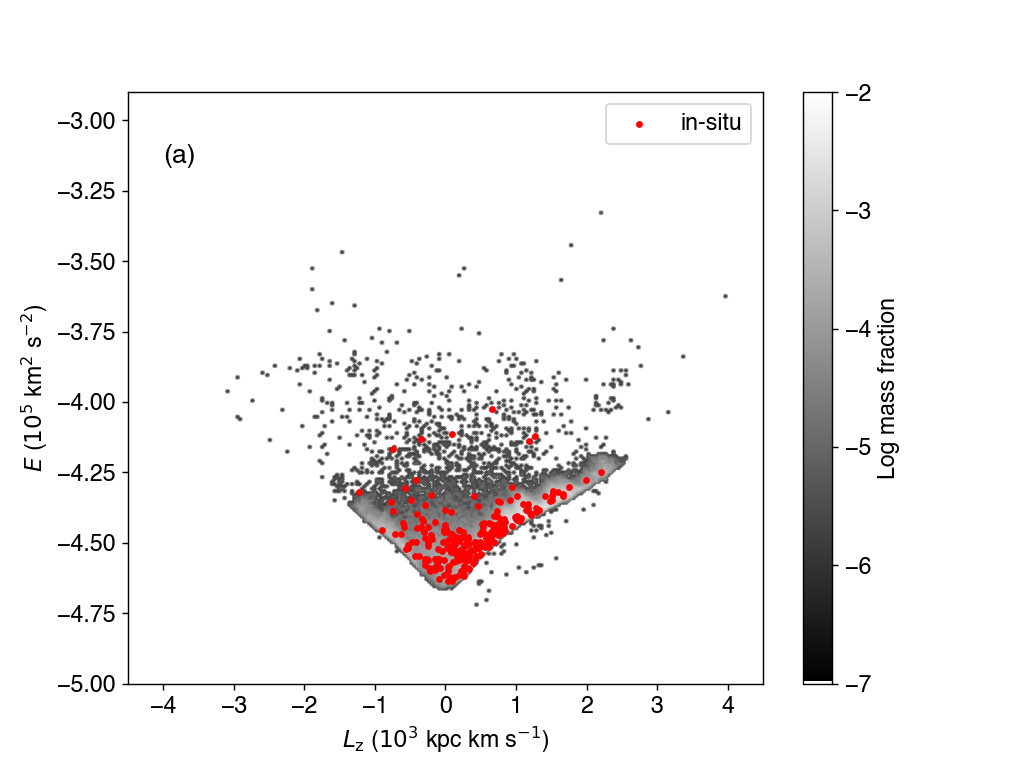}
	 \includegraphics[width=\columnwidth, bb = 0 550 550 0]{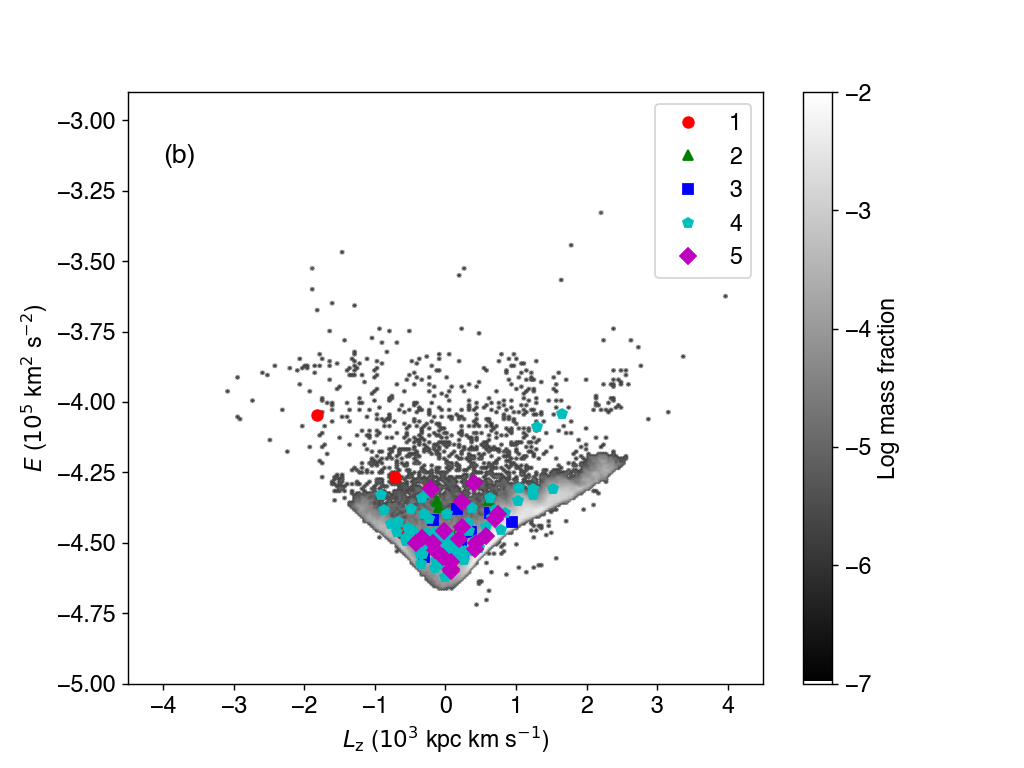}
     \includegraphics[width=\columnwidth, bb = 0 0 550 1000]{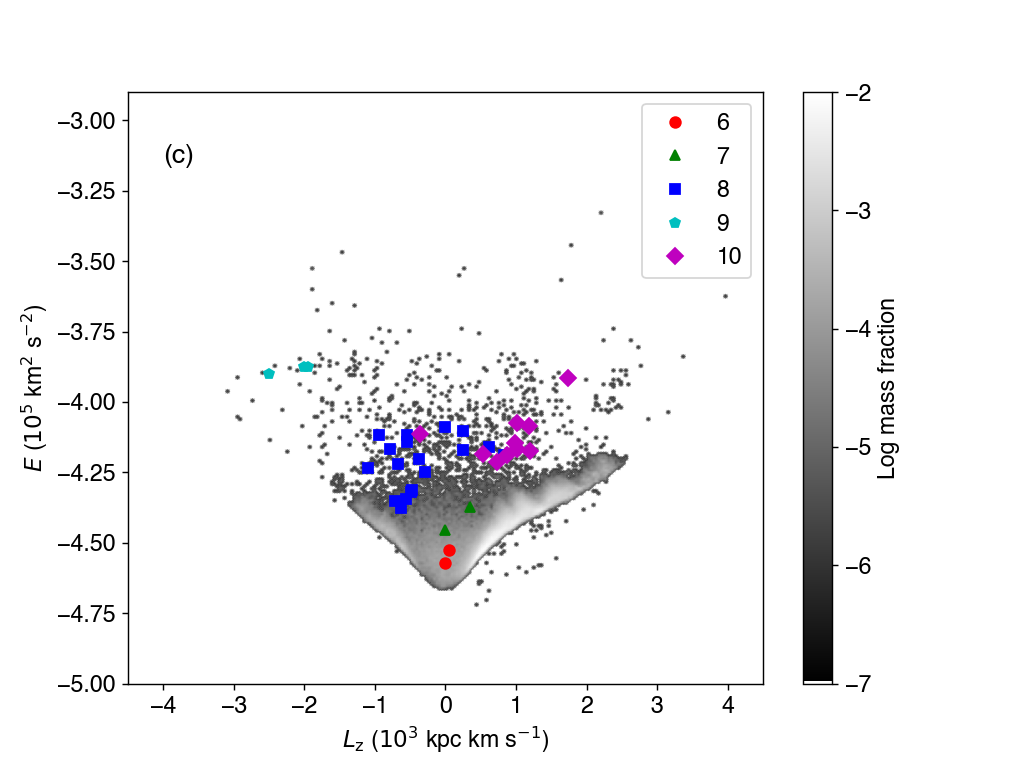}
     \includegraphics[width=\columnwidth, bb = 0 0 550 1000]{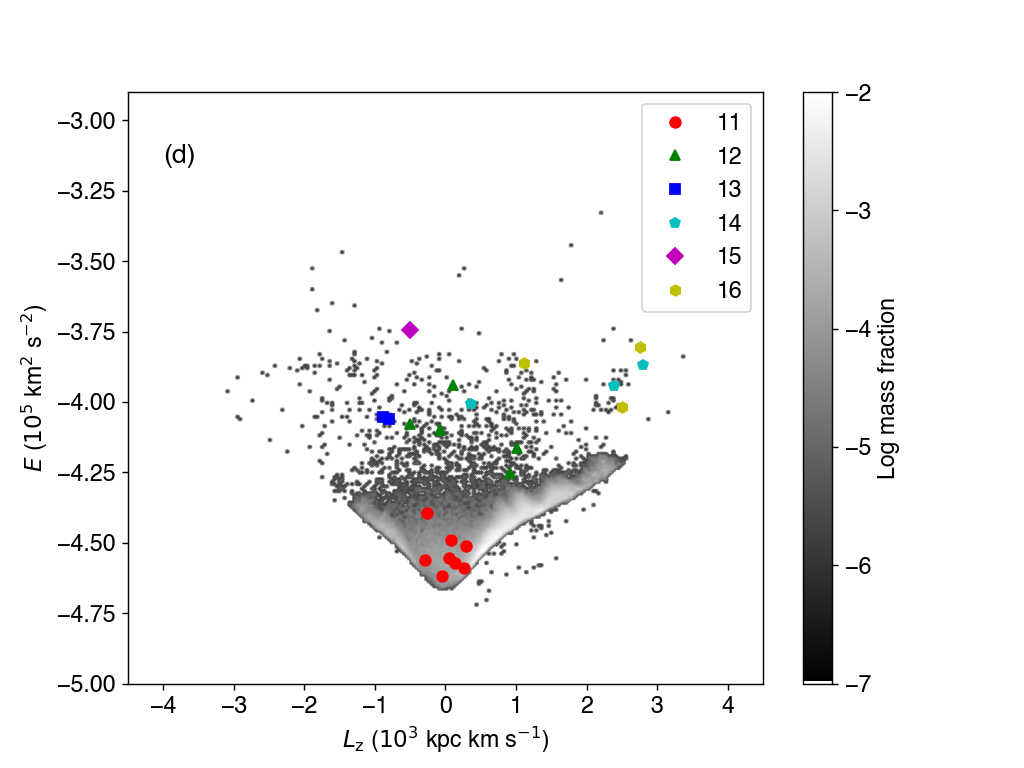}
    \caption{Distribution of star particles in the Lindblad diagram; total energy ($E$) versus the vertical angular momentum ($L_{\rm{z}}$) for (a) \textit{in situ} $r$-II stars, (b) stars formed in haloes 1 to 5, (c) 6 to 10 and (d) 11 to 16. The gray scale represents the mass fraction of all stars within 5 kpc from the `sun' and outside of 3 kpc from the Galactic Centre. The coloured {symbols} show $r$-II stars from the same haloes. The halo numbers are indicated in the legend.}
    \label{fig:ELz}
\end{figure*}

In order to extract information on the birth environments of $r$-II stars from the observations, it is necessary to clarify the chemo-dynamical properties of $r$-II stars. Integral of motions such as energy and angular momentum are conserved on long timescales and have been used to identify accreted components \citep[e.g.][]{2020ARA&A..58..205H}. Fig. \ref{fig:ELz} shows a Lindblad Diagram (the total specific energy, $E$, as a function of the orbital angular momentum, $L_{\rm{z}}$) for $r$-II stars. The total specific energy is the sum of the kinetic and potential energy per stellar mass. The kinetic energy is computed based on the velocity of the particle, and the potential energy is calculated using the distance and mass of all particles involved in the force calculation. The orbital angular momentum is defined to be $L_{\rm{z}}=RV_{\phi}$, where $R=\sqrt{x^2+y^2}$. Stars with low $E$ ($E\,<-4.5\,\times\,10^5\,$km$^2$\,s$^{-2}$) and high $L_{\rm{z}}$ ($L_{\rm{z}}\,>\,0.7\times\,10^3\,$kpc\,s$^{-2}$) are newly formed stars with ages $\lesssim$\,50 Myr around the solar neighbourhood. Since these stars are not observed in the MW and are not generally $r$-process enhanced, we exclude these stars from the discussion.

As shown in Fig. \ref{fig:ELz} (a), most \textit{in situ} $r$-II stars exhibit disc-like kinematics. The solar neighbourhood disc stars reside in the parabola of the $E$ vs $L_{\rm{z}}$ diagram with $L_{\rm{z}}\,>\,0$. These stars are formed in the Galactic disc's locally $r$-process-enhanced regions. \citet{shank2022c} find 30 $r$-II stars with disc-like kinematics in the RPA sample; such 
stars are candidates for stars formed by local inhomogeneities of $r$-process enhancement in the Galactic disc.

On the other hand, a few \textit{in situ} stars have retrograde orbits, high energy, or both. Most of these stars are formed in the primordial main progenitor halo before the disc formation. The $r$-II stars with the lowest binding energy ($E\,<-4.2\,\times\,10^5$ km$^2$ s$^{-2}$) are also old (ages with $\gtrsim$ 11 Gyr). These stars are associated with the main progenitor halo. The kinematics of these $r$-II stars are similar to the \textit{in situ} halo stars, as defined in \citet{2021ApJ...908..191C}. These authors pointed out that \textit{in situ} halo stars are located in the lowest binding energy or along a parabola in the $E$ vs $L_{\rm{z}}$ diagram. These results suggest that $r$-II stars with the lowest binding energy or located along the parabola of the Lindblad diagram could be formed in the main progenitor halo.

Fig. \ref{fig:ELz} (b)--(d) shows $E$ vs $L_{\rm{z}}$ for $r$-II stars formed in accreted components. As expected from \citet{2000MNRAS.319..657H}, stars formed in the same halo tend to reside in a similar region in their integrals of motion space. \citet{2018AJ....156..179R} found eight groups for $r$-II stars in the phase space by applying the clustering methods. \citet{shank2022c}
identify 38 CDTGs among $\sim$1800 RPE stars. The results of these studies indicate that $r$-II stars clustered in dynamical phase space could come from accreted components. According to Fig. \ref{fig:ELz}, some groups of $r$-II stars from the same halo are located in similar regions in the $E$ vs $L_{\rm{z}}$ diagram (e.g. haloes 8, 10 and 11). For example, the standard deviation around the cluster central value  ($\sigma_{\rm{d}} = \sqrt{\sigma_{Lz}^2+\sigma_{E}^2}$, where $\sigma_{Lz}$ and $\sigma_{E}$ are standard deviations of $L_{\rm{z}}$ and $E$, respectively) of halo 11 is $\sigma_{\rm{d}}\,=0.23$ while that of halo 0 is $\sigma_{\rm{d}}\,=0.57$. These results support the scenario that $r$-II stars clustered in their dynamical phase space come from accreted components.

The $r$-II stars from more massive haloes tend to be broadly distributed in the dynamical phase space. For example, haloes 4 and 5 have the peak stellar mass of 9.2$\,\times\,10^7\,$M$_{\sun}$ and 9.3$\,\times\,10^7\,$M$_{\sun}$, respectively. Stars in these haloes exist in $-1.0\,\times\,10^3$ kpc km s$^{-1}\,\lesssim L_{\rm{z}}\,\lesssim 2.0\,\times\,10^3$ kpc km s$^{-1}$ and $-4.6\,\times\,10^5$ km$^2$ s$^{-2}\,\lesssim E\lesssim -4.0\,\times\,10^5$ km$^2$ s$^{-2}$ (Fig. \ref{fig:ELz} (b)). In contrast, halo 11 has a peak stellar mass of 2.4$\,\times\,10^6\,$M$_{\sun}$. This halo is highly clustered in the phase space (Fig. \ref{fig:ELz} (d)), implying that $r$-II stars that are tightly clustered could originate from low-mass building blocks.

\begin{figure*}
	 \includegraphics[width=\columnwidth, bb = 0 350 400 0]{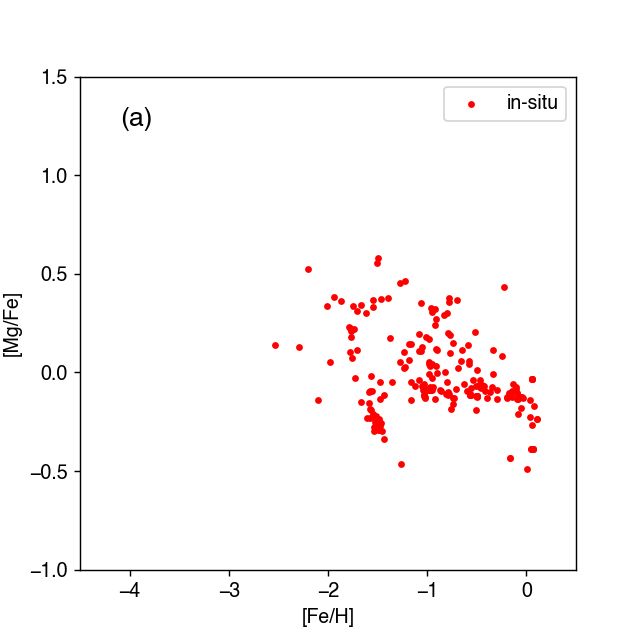}
	 \includegraphics[width=\columnwidth, bb = 0 350 400 0]{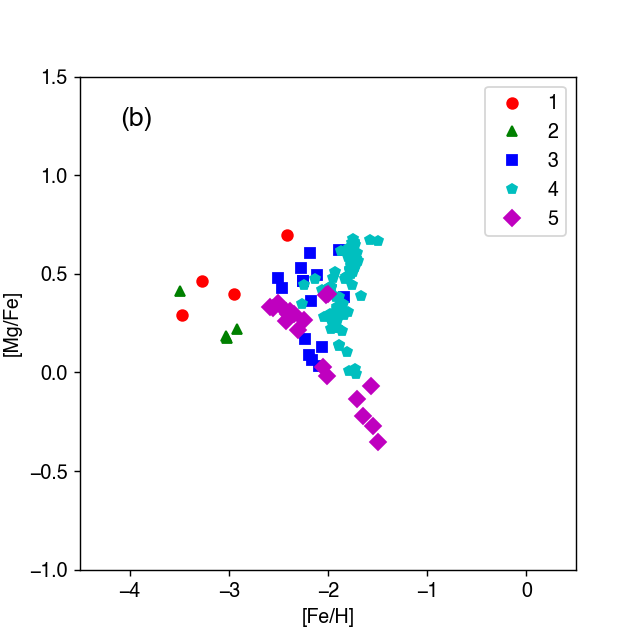}
     \includegraphics[width=\columnwidth, bb = 0 0 400 710]{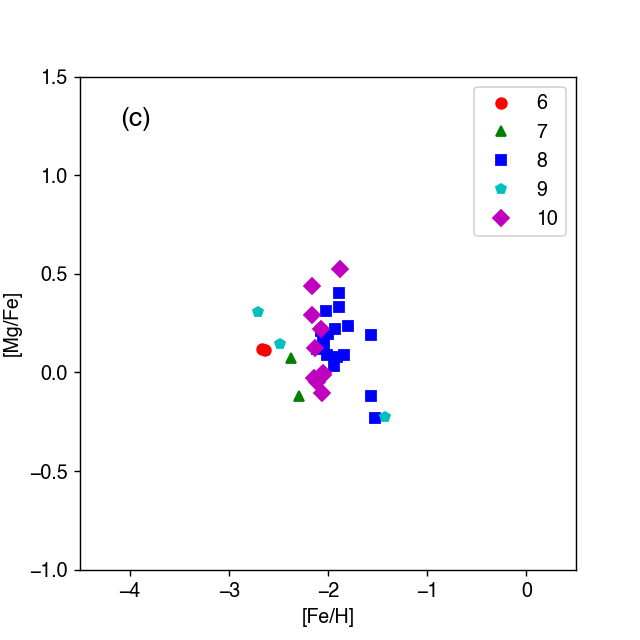}
     \includegraphics[width=\columnwidth, bb = 0 0 400 710]{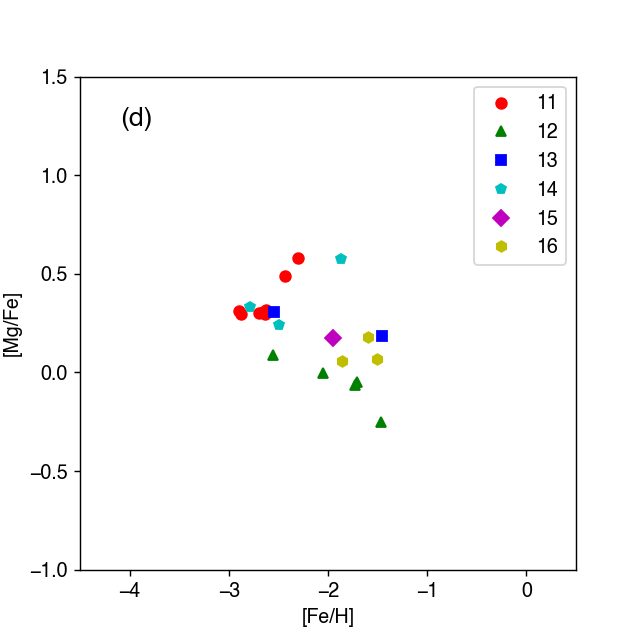}
    \caption{[Mg/Fe], as functions of [Fe/H], for (a) \textit{in situ} $r$-II stars, (b) stars formed in haloes 1 to 5, (c) 6 to 10 and (d) 11 to 16. Colour plots show $r$-II stars from the same haloes. The halo numbers are indicated in the legend.}
    \label{fig:MgFe_acc}
\end{figure*}

The $r$-II stars formed in haloes that were accreted earlier tend to have low energy. For instance, haloes 6 and 11, which reside in the lowest energy in the phase space, are examples of haloes that were accreted early in the simulation. From Table \ref{tab:group}, these haloes are accreted into the main halo at $t_{\rm{acc}}$ = 0.57 Gyr (halo 6) and 0.94 Gyr (halo 11). In contrast, stars from haloes 9 and 16 have high energy and much later accretion times ($t_{\rm{acc}}$ = 8.44 Gyr and 9.78 Gyr, respectively). Fig. \ref{fig:MgFe_acc} shows [Mg/Fe], as a function of [Fe/H], grouped by $r$-II stars from the same haloes. As shown in this figure, early accreted haloes (haloes 6 and 11) are characterized by low metallicity ([Fe/H]$\,<-2$) and high [Mg/Fe] ratios ([Mg/Fe]$\,>\,0$). In contrast, later accreted haloes (haloes 9 and 16) exhibit higher metallicity ([Fe/H]$\,>-2$) and lower [Mg/Fe] ratios ([Mg/Fe]$\,\sim 0$) compared to earlier accreted haloes, indicating that the latter haloes experienced contributions of iron from SNe Ia.

Some groups of $r$-II stars exhibit a decreasing trend of [Mg/Fe] ratios, typically seen in dwarf galaxies and accreted components. For \textit{in situ} stars, there is an overall decreasing trend of [Mg/Fe] for [Fe/H]$\,> -2$ (Fig. \ref{fig:MgFe_acc} (a)). Note that the trend is unclear because we adopt the short minimum delay time for SNe Ia ($t_{\rm{min}}$ = 40 Myr, Section \ref{sec:Code}). 
For accreted stars, some haloes (e.g. haloes 4 and 5) clearly show decreasing trends of [Mg/Fe] at [Fe/H]$\,\sim -2$. On the other hand, relatively low-mass haloes (e.g. haloes 1, 2, and 11) do not exhibit decreasing trends. These haloes halted star formation before SNe Ia started contributing to the Fe enrichment. SNe Ia begin to contribute to the Fe enrichment after 40 Myr from the first-star formation in each halo in this simulation.

The observed $r$-II stars generally have similar [Mg/Fe] ratios, as found in this study. \citet{2019NatAs...3..631X} describe the star LAMOST J112456.61+453531.3 with [Fe/H] = $-$1.27, [Eu/Fe] = +1.1. and [Mg/Fe] = $-$0.31 {\citep[see also, ][]{2019ApJ...874..148S}}. They argued that this star could come from a disrupted dwarf galaxy. In our simulation, there is a star with [Fe/H] = $-$1.49, [Eu/Fe] = +1.27 and [Mg/Fe] = $-$0.35 formed in halo 5. This star is formed at $t$ = 6.64 Gyr when the stellar mass of halo 5 is 4.10$\,\times\,10^7\,$M$_{\sun}$. We also find a similar \textit{in situ} halo star. This star has [Fe/H] = $-$1.26, [Eu/Fe] = +1.75 and [Mg/Fe] = $-$0.46, formed at $t$ = 6.68 Gyr when the stellar mass of the halo is 9.20$\,\times\,10^7\,$M$_{\sun}$. These results suggest that $r$-II stars with high metallicity and low [Mg/Fe] ratios can arise from the $r$-process-enhanced gas clumps formed in relatively massive building blocks.

Many of the $r$-II stars with [Fe/H]$\,<-2$ exhibit [Mg/Fe]$\,>0$. According to Fig. \ref{fig:MgFe_acc}, most $r$-II stars have [Mg/Fe] > 0. For example, \citet{shank2022c} find that most $r$-II stars in the RPA sample have [Mg/Fe]$\,>0$. 
These results suggest that $r$-II stars with [Fe/H]$\,<-2$ are mainly formed in haloes without significant contributions from SNe Ia, i.e. those in an early stage of their chemical evolution.

\begin{figure*}
	 \includegraphics[width=\columnwidth, bb = 0 350 400 0]{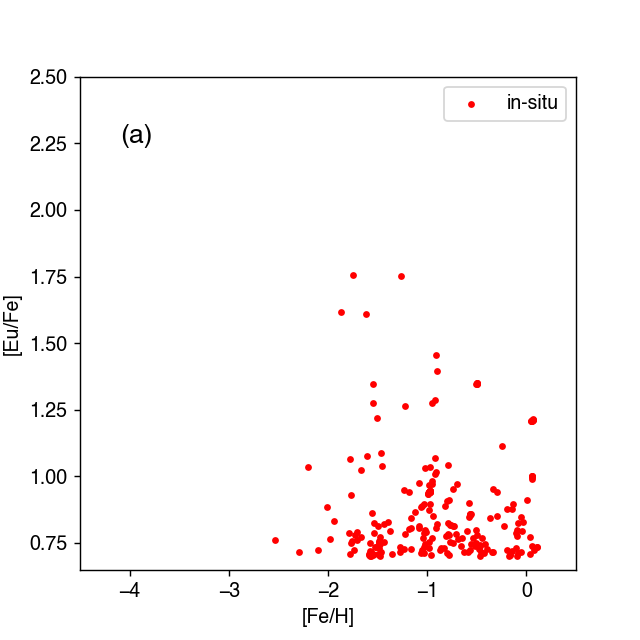}
	 \includegraphics[width=\columnwidth, bb = 0 350 400 0]{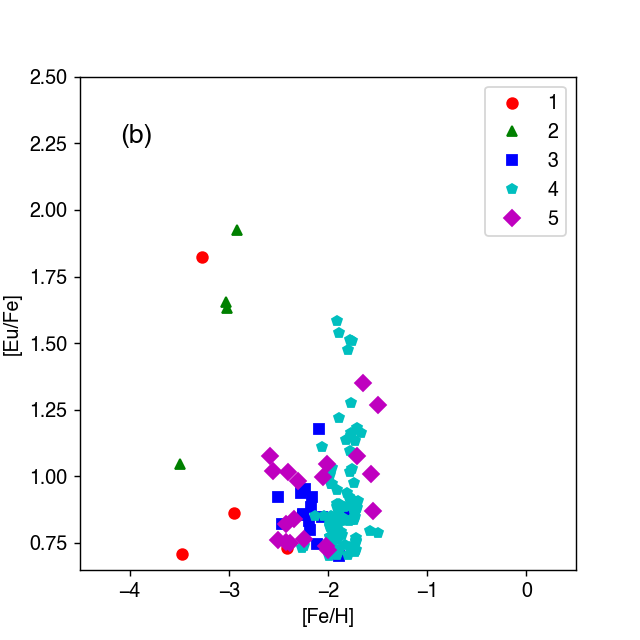}
     \includegraphics[width=\columnwidth, bb = 0 0 400 710]{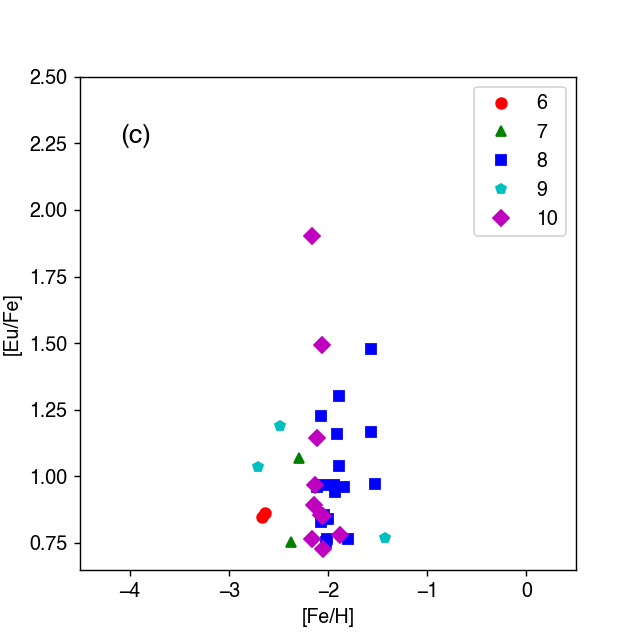}
     \includegraphics[width=\columnwidth, bb = 0 0 400 710]{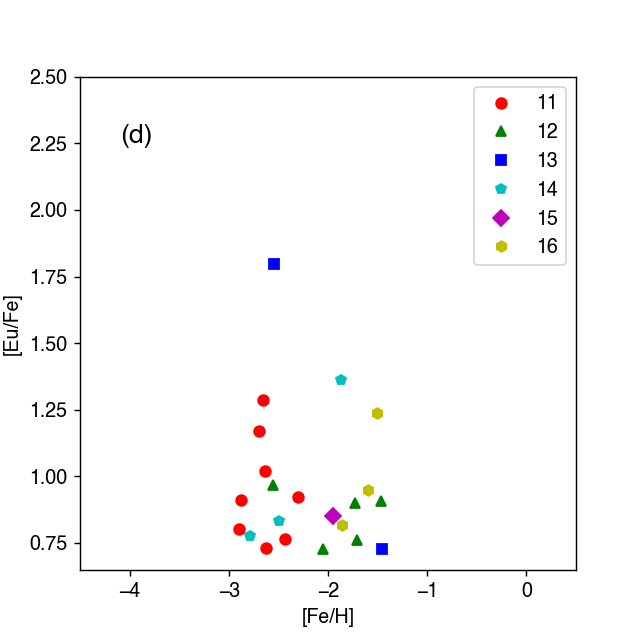}
    \caption{Same as Fig. \ref{fig:MgFe_acc}, but for [Eu/Fe] as a function of [Fe/H].}
    \label{fig:EuFe_acc}
\end{figure*}

The $r$-II stars in each accreted component exhibit low dispersions in [Fe/H] and larger variations of their [Eu/Fe] ratios. For example, Fig. \ref{fig:EuFe_acc} depicts [Eu/Fe], as a function of [Fe/H], for \textit{in situ} $r$-II stars and those in each halo. This figure shows that $r$-II stars in the same accreted component have low scatter in [Fe/H], while \textit{in situ} $r$-II stars are seen over a wide metallicity range. All of the accreted components with more than three stars have standard deviations of [Fe/H] less than 0.5 dex. This result is consistent with observed $r$-II stars in CDTGs \citep{2018AJ....156..179R, 2020ApJsubGudin}. \citet{shank2022c} show that the standard deviations of [Fe/H] in CDTGs with more than five stars are less than 0.33 dex, and those in [Eu/Fe] are less than 0.16 dex. {Note that this estimate includes $r$-I stars. Scatters of [Eu/Fe] in accreted components, including $r$-I stars, will be examined in our forthcoming paper.} The short timescale for $r$-II star formation results in the low scatter in [Fe/H]. The longest time separation of the first and the last $r$-II star formation is 620 Myr in halo 12. Most haloes form $r$-II stars within 100 Myr of one another.

The dispersion of [Eu/Fe] ratios seen in accreted components could explain the $r$-II star found in the Indus stellar stream. \citet{2021ApJ...915..103H} reported that a star Indus\_13 shows [Eu/Fe] = +1.81, while other Indus member stars have [Eu/Fe] $\approx$ +0.5. Halo 1 in our simulation shows a similar trend. In this halo, the oldest and most RPE star has [Eu/Fe] = +1.82, but other stars have +0.71$\,\leq\,$[Eu/Fe]$\,\leq\,$+0.86. The lack of stars between these values is because we only select solar neighbourhood stars. The scatter seen in accreted components can be caused by these haloes' inhomogeneous [Eu/Fe] distribution, as even if their average [Eu/Fe] is enhanced, there are large scatters of the gas-phase [Eu/Fe] ratios from [Eu/Fe]$\,<-1$ to [Eu/Fe]$\,>\,+2$.

\subsection{Requisite mass resolution}\label{sec:resolution}
{Here we discuss the resolution required to resolve the scatter of [Eu/Fe] abundance ratios caused by the building blocks and in the \textit{in situ} component.  In order to accurately compute the chemo-dynamical properties of the building blocks, the dark matter, gas, and stellar components need to be resolved. The typical mass and the virial radius of dark matter haloes that form UFD-mass galaxies is $\sim$10$^9\,$M$_{\sun}$ and $\sim$10 kpc, respectively \citep[e.g.][]{2019MNRAS.490.4447W}. In such haloes, most stars are formed within $\sim$1 kpc. This means that it is necessary to resolve the inner structures of haloes down to at least 10 per cent of the virial radius. According to \citet{2003MNRAS.338...14P}, $\sim$10$^3$ particles need to be contained within 10 per cent of the virial radius to estimate the density profile of haloes robustly (see their figure 14). In this case, $\sim$10$^4$ dark matter particles should be contained within the virial radius. Therefore, simulations with $m_{\rm{DM}}\lesssim 10^5\,$M$_{\sun}$ in a zoomed-in region can resolve haloes that are able to form UFDs.}

{Gas clumps of $\sim$10$^6\,$M$_{\sun}$ need to be resolved to discuss the formation of RPE stars. \citet{2016Natur.531..610J} estimated that an NSM could explain the value of [Eu/H] seen in Reticulum II if the ejecta from the NSM is diluted by a gas cloud with $\sim$10$^6\,$M$_{\sun}$. SPH simulations evaluate physical quantities using $\sim$100 nearest neighbour particles, thus the smallest gas-mass scale for which one can evaluate physical quantities is $\sim$100 $m_{\rm{gas}}$. Therefore, simulations with $m_{\rm{gas}}\,\lesssim\,10^4\,$M$_{\sun}$ can resolve the formation of $r$-II stars in both the accreted and \textit{in situ} components.}

{For the stellar components, recent very high-resolution simulations of MW-like galaxies have shown that the size and velocity dispersions of satellite galaxies with more than 10 star particles successfully reproduce the size and the velocity dispersion of the observed UFDs \citep{2021ApJ...906...96A, 2021MNRAS.507.4953G}. This result indicates that we can obtain reliable stellar properties for galaxies with $\sim$10$^5\,$M$_{\sun}$ in simulations with star-particle masses of $m_{*}\lesssim\,10^4\,$M$_{\sun}$. As shown in Table \ref{tab:model}, all of these requirements are satisfied by our simulation. We note that these are minimum requirements, and further finer resolutions improve analyses. Thanks to this high resolution, our simulation can resolve the scatter of [Eu/Fe] in each RPE star-forming region and compare the kinematics of $r$-II stars with observed CDTGs.}

\subsection{Comparison to previous studies}\label{sec:comparison}
{As discussed in Section \ref{sec:sim}, some previous studies have considered the enrichment of $r$-process elements in their simulations of MW-like galaxies \citep{2015ApJ...807..115S, 2015MNRAS.447..140V, 2020MNRAS.494.4867V, 2022MNRAS.512.5258V, 2018MNRAS.477.1206N, 2019MNRAS.483.5123H}. Compared with these, our simulation has the highest stellar-mass resolution. Table \ref{tab:sim} compares the mass-resolution and metal-mixing models of previous simulations. The mass resolution affects the star-to-star scatter of [Eu/Fe] at low metallicity. All simulations with $m_{\rm{gas}}\,\lesssim\,10^4\,$M$_{\sun}$ produce such scatter. \citet{2015MNRAS.447..140V} have shown that their low-resolution model underestimates the observed scatter. \citet{2019MNRAS.483.5123H} could not reproduce the [Eu/Fe] scatter by their NSM models. Since the number of stars with [Fe/H]$\,<\,-$2.5 is much lower than for more metal-rich stars, and they tend to form in accreted components, it is difficult to correctly follow the formation of RPE stars in simulations with mass resolution lower than the requisite resolution ($m_{\rm{DM}}\,\lesssim\,10^5\,$M$_{\sun}$, $m_{\rm{gas}}\,\lesssim\,10^4\,$M$_{\sun}$, $m_{*}\,\lesssim\,10^4\,$M$_{\sun}$), as discussed in Section \ref{sec:resolution}.}

\begin{table*}
	\centering
	\caption{{Mass resolution of this work and previous studies. From left to right, the columns show the name of the simulation, the mass of one dark matter particle ($m_{\rm{DM}}$), the initial mass of one gas particle or cell ($m_{\rm{gas}}$), the initial mass of one star particle ($m_{\rm{*}}$), the final simulation redshift ($z_{\rm{final}}$), metal-mixing model, and references. Acronyms used for the 
	metal-mixing models are HS: turbulence-induced metal mixing model \citep{2010MNRAS.407.1581S,2017ApJ...838L..23H,2017AJ....153...85S}, SF: metal mixing in star-forming regions \citep{2015ApJ...807..115S, 2015ApJ...814...41H, 2017MNRAS.466.2474H}, and MM: metal mixing computed together with mass transfer among moving mesh \citep{2010MNRAS.401..791S}.}\label{tab:sim}}
	\begin{tabular}{lllllll}
		\hline
		Simulations &
		$m_{\rm{DM}}$ & $m_{\rm{gas}}$ &
		 $m_{*}$ & $z_{\rm{final}}$& Metal mixing & References
 	    \\
		 & (M$_{\sun}$) &  (M$_{\sun}$) &
		  (M$_{\sun}$) & &
		 \\		 
		\hline
		This work &
		7.2 $\times$ 10$^{4}$  & 
	    1.3 $\times$ 10$^{4}$ &
	    4.5 $\times$ 10$^{3}$ &
	    0&
	    HS
	    &
	    -
		\\
	    Eris &
		9.8 $\times$ 10$^{4}$  & 
	    2.0 $\times$ 10$^{4}$ &
	    6.0 $\times$ 10$^{3}$ &
	    0 &
	    SF
	    &
	    \citet{2015ApJ...807..115S}
	    \\
	    FIRE (Low) &
		2.3 $\times$ 10$^{6}$  & 
	    4.5 $\times$ 10$^{5}$ &
	    4.5 $\times$ 10$^{5}$  &
	    0 &
	    - &
	    \citet{2015MNRAS.447..140V}
	    \\
	    FIRE (Fiducial) &
		2.8 $\times$ 10$^{5}$  & 
	    5.7 $\times$ 10$^{4}$ &
	    5.7 $\times$ 10$^{4}$  &
	    0 &
	    - &
	    \citet{2015MNRAS.447..140V}
	    \\
	    FIRE (High) &
		3.5 $\times$ 10$^{4}$  & 
	    7.1 $\times$ 10$^{3}$ &
	    7.1 $\times$ 10$^{3}$  &
	    2.4 &
	    - &
	    \citet{2015MNRAS.447..140V}
	    \\
	    IllustrisTNG100-1 &
		7.5 $\times$ 10$^{6}$  & 
	    1.4 $\times$ 10$^{6}$ &
	    1.4 $\times$ 10$^{6}$ &
	    0 &
	    MM &
	    \citet{2018MNRAS.477.1206N}
	    \\
	    IllustrisTNG100-2 &
		6.0 $\times$ 10$^{7}$  & 
	    1.1 $\times$ 10$^{7}$ &
	    1.1 $\times$ 10$^{7}$ &
	    0 &
	    MM &
	    \citet{2018MNRAS.477.1206N}
	    \\
	    Aquila (G3-CK) &
		2.2 $\times$ 10$^{7}$  & 
	    3.5 $\times$ 10$^{6}$ &
	    1.8 $\times$ 10$^{6}$ &
	    0 &
	    - &
	    \citet{2019MNRAS.483.5123H}
	    \\
	    Auriga &
		4.3 $\times$ 10$^{4}$  & 
	    8.1 $\times$ 10$^{3}$ &
	    8.1 $\times$ 10$^{3}$ &
	    0 &
	    MM &
	    \citet{2020MNRAS.494.4867V}
	    \\	    
	    Auriga &
		3.6 $\times$ 10$^{4}$  & 
	    6.7 $\times$ 10$^{3}$ &
	    6.7 $\times$ 10$^{3}$ &
	    0 &
	    MM &
	    \citet{2022MNRAS.512.5258V}
	    \\	 
		\hline
	\end{tabular}
\end{table*}

{Metal mixing also plays a vital role in correctly computing the enrichment of $r$-process elements. \citet{2015ApJ...807..115S} improved the agreement between the predicted and observed scatter of [Eu/Fe] abundance ratios when they implemented a model for metal mixing in the star-forming region. \citet{2017ApJ...838L..23H} found that models without metal mixing overproduce RPE stars in dSphs. Thus, they calibrated the diffusion coefficient for metal mixing to suppress RPE star formation in dSphs. This study uses their value to compute the fraction of $r$-II stars in low-metallicity environments.}

{Both the simulations of this study and \citet{2020MNRAS.494.4867V} predict that very metal-poor stars are formed in the early phase of galaxy formation. \citet{2020MNRAS.494.4867V} reported that the median formation redshifts were $z$ = 4.7 (12.6 Gyr ago, [Fe/H]$\,<\,-$2) and $z$ = 6.2 (12.9 Gyr ago, [Fe/H]$\,<\,-$3). They also mentioned that the ages of outliers (in [$r$-process/Fe]) are older than the median in the same metallicity bin. Our simulation also finds that very and extremely metal-poor stars are formed at high-$z$: $z$ = 7.530 (13.18 Gyr ago, [Fe/H]$\,<\,-$2) and $z$ = 7.844 (13.22 Gyr ago, [Fe/H]$\,<\,-$3). We also confirm that the median age of $r$-II stars is older than all-stars: $z$ = 7.836 (13.22 Gyr ago, [Fe/H]$\,<\,-$2) and $z$ = 10.44 (13.41 Gyr ago, [Fe/H]$\,<\,-$3). The older ages in our study compared to \citet{2020MNRAS.494.4867V} could be due to differences in the efficiency of metal mixing and initial conditions. Nevertheless, these results suggest that the median ages of very metal-poor stars are over 10 Gyr, and those of $r$-II stars are even older.}

{While this study and \citet{2020MNRAS.494.4867V} predict a higher fraction of RPE stars formed in accreted components (Fig. \ref{fig:f_acc} (a)) than that of non-RPE stars,  \citet{2018MNRAS.477.1206N} could not find any correlation with the Eu abundance and assembly history. This discrepancy could come from the difference in the resolution. The average mass gas cell in \citet{2018MNRAS.477.1206N} is $1.4\times10^6\,$M$_{\sun}$. With this resolution, it is not possible to resolve UFD-like building blocks, which are key to forming $r$-II stars (Section \ref{sec:form}). Most of the $r$-II stars in their simulation would be formed in local inhomogeneities within the disc or haloes.}

\subsection{Caveats}\label{sec:caveats}
{For [Fe/H] $\gtrsim -1$, the simulated [Eu/Fe] abundance ratios in Fig. \ref{fig:EuFe} are constant towards higher metallicity, contrary to the observations, which exhibit a decreasing trend of [Eu/Fe]. This result is because we adopt a power-law index of $-1$ for the delay-time distributions of NSMs, which is the same power-law index adopted for SNe Ia. The effects of increasing [Eu/Fe] ratios by NSMs and the expected decrease due to SNe Ia offset each other under this assumption \citep{2018IJMPD..2742005H}.  As a result, the fraction of $r$-II stars at high metallicity is expected to be over-estimated. This problem could be resolved by considering $r$-process production from short time-scale events \citep{2019ApJ...875..106C}, a longer minimum delay time for SNe Ia \citep{2021MNRASW}, natal kicks, and inside-out disc evolution \citep{2020ApJ...902L..34B}, possibly in combination. These effects start to dominate for [Fe/H]$\,\gtrsim -1$. Therefore, the lack of modelling of these effects does not largely affect the [Eu/Fe] distribution with [Fe/H]$\,\lesssim -1$.}

{This simulation tends to produce more $r$-II stars in this metallicity range compare to observations. There are two reasons for this tendency: (1) Since this simulation assumes the similar delay-time distributions for SNe Ia and NSMs, the average [Eu/Fe] ratios are higher than the observations for [Fe/H]$\,>\,-1$, as discussed above. Due to the high average [Eu/Fe] abundance ratios, the [Eu/Fe] in $r$-process-enhanced gas are not significantly diluted by the surrounding gas; and (2) the value of the diffusion coefficient assumed in this simulation could be insufficient. In this study, we adopt a lower value of the scaling factor for the metal diffusion ($C_{\rm{d}}\,=0.01$). However, \citet{2017ApJ...838L..23H} could not constrain the value of $C_{\rm{d}}$ larger than 0.01. Adopting a higher value of $C_{\rm{d}}$ would suppress the formation of $r$-II stars.}

{The existence and persistence of the local inhomogeneity caused by an NSM (Fig. \ref{fig:EuFegas}) would depend on the resolution and the efficiency of the dilution of metals. As discussed in Section \ref{sec:form}, the mass of the gas clump is typically $\sim$10$^6\,$M$_{\sun}$, which is the smallest mass scale to evaluate the physical quantities in our SPH simulation. The timescale of the dilution is also affected by the metal-diffusion coefficient \citep{2017ApJ...838L..23H}. This should be tested in future higher-resolution simulations.}

{The dispersions seen in Fig. \ref{fig:EuFe_acc} could also be due to insufficient metal mixing in our simulation. Since metal mixing is caused by turbulence in the interstellar medium, which is not resolved in the simulation, the efficiency of metal mixing on a galactic scale is unknown. \citet{2017ApJ...838L..23H} have shown that the scatter of [Eu/Fe] can be artificially caused in SPH simulations with low efficiencies of metal mixing. Although we adopt the recommended value from their study, which is consistent with other estimates on different scales \citep[e.g.][]{1987JCoPh..71..343H, 2000AnRFM..32....1M}, there is still the possibility that our simulation does not have sufficient efficiency to thoroughly mix metals. Determining the dispersions of [Eu/Fe] from high-resolution spectroscopic observations in accreted components and dwarf galaxies could help empirically resolve this issue.}

In this study, we have concluded that $r$-II stars can be formed in low-mass building blocks enhanced in $r$-process elements or from locally enhanced gas clumps in relatively massive building-block galaxies and the main halo. The birthplaces of $r$-II stars depend on metallicity. For [Fe/H]$\,<-2$, most of the $r$-II stars are formed in accreted components (Fig. \ref{fig:f_acc} (a)). Especially for [Fe/H]$\,<-2.5$, $r$-II stars are formed in low-mass dwarf galaxies with stellar mass $\sim$10$^5\,$M$_{\sun}$ when their gas was enhanced in $r$-process elements (haloes 1, 2, 6, 9, 11, 13 and 14). NSMs are allowed to occur at low metallicity in such low-mass systems because of their slow chemical evolution. For higher metallicity, $r$-II stars are formed in local inhomogeneities of $r$-process abundances in massive building blocks or the main halo. Although the local enhancement of $r$-process abundances is erased within a relatively short timescale ($\sim$10 Myr), $r$-II stars can be formed if the system is sufficiently massive to host NSMs that lead to star formation in the $r$-process-enhanced gas clumps \citep{2016Natur.531..610J, 2016AJ....151...82R, 2018ApJ...865...87O, 2019ApJ...871..247B, 2021MNRASW}. Simultaneously, it is possible to explain the existence of high-metallicity RPE stars by star formation in local inhomogeneities of the gas-phase $r$-process abundances.

\section{Conclusions}\label{sec:conclusion}

In this paper, we perform a cosmological zoom-in simulation of an MW-like galaxy to investigate the birth environments and chemo-dynamical properties of $r$-II stars, one of the first with sufficient mass ($m_{\rm{gas}}\,=\,1.3\,\times\,10^4\,$M$_{\sun}$) and time resolution ($\sim$10 Myr) to accomplish this purpose. This simulation produces star-to-star dispersions of [Eu/Fe] ratios at [Fe/H]$\,<-2$ (Fig. \ref{fig:EuFe}), similar to the observations, thanks to appropriate modelling of NSMs and metal diffusion. We confirm that most $r$-II stars with [Fe/H]$\,<-1$ exhibit halo-like orbits reported in previous observations \citep{2018AJ....156..179R, 2020ApJsubGudin, shank2022c}.


We find that over 90 per cent of $r$-II stars are formed in the early epochs of the galaxy formation ($t\,\lesssim 4\,$Gyr), suggesting that the majority of $r$-II stars are formed in small building blocks and accreted into the main halo later (see Fig. \ref{fig:FeHTime}). 
We also find an increasing trend in the fraction of accreted components towards low metallicity (see Fig. \ref{fig:f_acc}) and that the accreted fraction of $r$-II stars is higher than that of non-RPE stars. These results indicate that the majority of $r$-II stars were formed in disrupted dwarf galaxies over 10 Gyr ago.

The $r$-II stars are formed in gas clumps enhanced in $r$-process elements. For [Fe/H]$\,< -2.5$, $r$-II stars are formed in dwarf galaxies with $M_*\lesssim 10^5\,$M$_{\sun}$. Because of the small gas mass ($M_{\rm{gas}}\,\lesssim 10^7\,$M$_{\sun}$) of these galaxies, the average gas-phase [Eu/Fe] is enhanced once an NSM occurs. In such a system, most stars are formed as RPE stars. We find an increasing trend in the fraction of $r$-II stars towards haloes with lower stellar mass. For [Fe/H]$\,>-2.5$, $r$-II stars are mainly formed in locally $r$-process-enhanced gas clumps in relatively massive accreted components or the main halo. This formation channel could account for the $r$-II stars found in classical dSphs \citep[e.g.][]{2021ApJ...912..157R} and the MW star with [Fe/H]$\,=-1.27$ \citep{2019NatAs...3..631X}. 

The chemo-dynamical properties of $r$-II stars also provide valuable information on their origin. The $r$-II stars from accreted components tend to be clustered in their dynamical phase space, while the majority of $r$-II stars formed \textit{in situ} have disc-like kinematics. We find that some dynamical groups of $r$-II stars exhibit decreasing [Mg/Fe] towards higher [Fe/H], similar to present dwarf galaxies. We also find that $r$-II stars in the same halo exhibit low dispersions of [Fe/H] (since $r$-II stars are formed on relatively short timescales, $\sim$100\,Myr) and somewhat larger dispersions of [Eu/Fe] in each halo, as also seen from the observations. Furthermore, the fraction of halo $r$-II stars found in our simulation is commensurate with observations from the RPA, and crucially, the distribution of the predicted [Eu/Fe] for halo $r$-II stars well matches that observed. The dispersion of [Eu/Fe] in each halo could also explain the $r$-II star in the Indus stellar stream  \citep{2021ApJ...915..103H}. These results clearly indicate that observations of $r$-process elements, together with $\alpha$-elements and kinematics, help distinguish metal-poor stars' formation environments.

Efforts to identify and analyse RPE stars and perform refined high-resolution galaxy formation simulations will greatly improve our understanding of the MW's formation in the earliest epochs. As shown in this study, most $r$-II stars are formed in the accreted components in the early Universe. Detailed comparisons with the chemo-dynamical properties of RPE stars and galaxy-formation simulations can constrain the evolutionary histories of the building blocks of the MW and the astrophysical sites responsible for the production of the heaviest elements. Observationally, the RPA and \textit{Gaia} are greatly increasing the number and accuracy of the derived chemo-dynamical properties of RPE stars. Eventually, individual stellar abundances and kinematics will be used to compare with future galaxy-formation simulations resolving individual stars \citep[e.g.][]{2017MNRAS.471.2151H, 2019MNRAS.483.3363H, 2020ApJ...891....2L, 2021MNRAS.501.5597G,2022MNRAS.513.1372G, 2021PASJ...73.1036H, 2021PASJ...73.1057F, 2021PASJ...73.1074F}. These efforts will open up a new window to reveal galaxy formation and nucleosynthesis in the early Universe.

\section*{Acknowledgements}
{We appreciate the anonymous referee for providing valuable suggestions to improve the manuscript. We also thank Ian U. Roederer for his careful reading of the draft.} We are grateful for fruitful discussions with Alexander P. Ji and Ting S. Li. This work was supported in part by JSPS KAKENHI Grant Numbers JP21J00153, JP20K14532, JP21H04499, JP21K03614, JP22H01259, JP21H05448, JP18H05437, JP21H00055, JP21H04500, 
JP21K03633, JP20H05861, JP21H04496, JP19H01931, JP20H01625, JP21K11930, MEXT as `Program for Promoting Researches on the Supercomputer Fugaku' (Toward a unified view of the Universe: from large scale structures to planets, Grant No. JPMXP1020200109){, JICFuS}, grants PHY 14-30152; Physics Frontier Center/JINA Center for the Evolution of the Elements (JINA-CEE), and OISE-1927130: The International Research Network for Nuclear Astrophysics (IReNA), awarded by the US National Science Foundation. Numerical computations and analysis were carried out on Cray XC50 and computers at the Center for Computational Astrophysics, National Astronomical Observatory of Japan and the  Yukawa Institute Computer Facility. This research has made use of NASA's Astrophysics Data System {and Astropy,\footnote{http://www.astropy.org} a community-developed core Python package for Astronomy \citep{2013A&A...558A..33A, 2018AJ....156..123A}}.
\section*{Data Availability}
Data will be available with a reasonable request to the corresponding author. \textsc{celib} is available at {\tt https://bitbucket.org/tsaitoh/celib}.
\bibliographystyle{mnras}
\bibliography{sampleNotes}

\bsp	
\label{lastpage}
\end{document}